\documentclass[galaxies,review,accept,pdftex,moreauthors]{Definitions/mdpi} 

\makeatletter
\newcommand{\customlabel}[2]{%
   \protected@write \@auxout {}{\string \newlabel {#1}{{#2}{\thepage}{#2}{#1}{}} }\hypertarget{#1}{}%
}
\makeatother

\firstpage{1} 
\pubvolume{11}
\issuenum{11}
\articlenumber{87}
\pubyear{2023}
\copyrightyear{2023}
\externaleditor{Academic Editor: Jose L. Gómez}
\datereceived{9 June 2023} 
\daterevised{4 July 2023} 
\dateaccepted{17 July 2023} 
\datepublished{19 July 2023} 
\hreflink{https://doi.org/10.3390/\linebreak galaxies11040087} 

\Title{Dynamics of Powerful Radio Galaxies}

\TitleCitation{Dynamics of Powerful Radio \linebreak Galaxies}

\Author{{Ross J. Turner} 
	*\orcidA{} and Stanislav {S.} 
	Shabala \orcidB{}}

\AuthorNames{Ross J. Turner and Stanislav S. Shabala}

\AuthorCitation{{Turner, R.J.}; Shabala, S.S.}

\address[1]{School of Natural Sciences, University of Tasmania, Private Bag 37, Hobart 7001, Australia
}

\corres{\hangafter=1 \hangindent=1.05em \hspace{-0.82em}Correspondence: ross.turner@utas.edu.au}

\abstract{Analytical models describing the dynamics of lobed radio sources are essential for interpretation of the tens of millions of radio sources that will be observed by the Square Kilometre Array and pathfinder instruments. We propose that historical models can be grouped into two classes in which the forward expansion of the radio source is driven by either the jet momentum flux or lobe internal pressure. The most recent generation of analytical models combines these limiting cases for a more comprehensive description. 
	We extend the mathematical formalism of historical models to describe source expansion in non-uniform environments, and directly compare different model classes with each other and with hydrodynamic numerical simulations. 
	We quantify differences in predicted observable characteristics for lobed radio sources due to the different model assumptions for their dynamics. We have made our code for the historical models analysed in this review openly available to the~community.
}


\begin{document}
	

	\section{Introduction}
	
	%
	%
	
	The first extragalactic radio sources were identified over seven decades ago by Bolton in the late-1940s \citep{Bolton+1949}. Shortly after, in 1953, the first resolved image was captured of Cygnus A, now known as the archetypal ``classical double'' \citep{Jennison+1953}. This breakthrough was followed by the pioneering efforts of radio survey groups in Australia \citep{Mills+1957} and the United Kingdom~\citep{Edge+1959}, which conducted the first large-scale radio surveys (for a comprehensive review, refer to \citep{Norris+2017}). The first quasar, 3C273, was discovered a few years later in 1963 ({see} 
	\citep{Hazard+2018,Hazard+1963,Schmidt+1963} for a historical review). These pivotal developments laid the foundation for observational studies of radio galaxies. Subsequently, building upon Lynden-Bell's (\citep{Lynden-Bell+1969}, {1969}) 
	proposal that black holes are responsible for the extreme luminosities observed in quasars, the 1970s saw the development of the first models describing the dynamical evolution of radio~galaxies.
	
	While varying in specific details, the majority of models in the literature share a similar overarching framework. These models in general consider two initially conical jets composed of particles that have been accelerated to relativistic velocities. The interaction between the jets and the intracluster medium (ICM) surrounding their host galaxy determines the subsequent evolution. 
	Jets which retain sufficient forward ram pressure during their initial propagation phase (typically on galaxy scales) will be collimated by pressure from the ambient medium, or more likely, a build-up of plasma shed by the jet in the early stages of lobe formation \citep{Alexander+2006, Krause+2012}. Regardless, each collimated jet leads to the formation of a Mach disk (observable as a hotspot) and, as overpressured jet material flows back towards the equatorial plane, the inflation of a plasma lobe observable through synchrotron radiation; such objects are generally classified to have a Fanaroff and Riley {\cite{Fanaroff+1974}} Type-II (FR-II) lobe morphology.
	
	\textls[10]{On the other hand, jets which suffer substantial entrainment (e.g., from stellar winds,~\citep{Perucho+2014, Wykes+2015} or the interstellar medium, \citep{Bicknell+1995}) will slow down to transsonic speeds (with respect to the internal lobe sound speed, of order 0.1c) and be disrupted. In this scenario, the jet momentum thrust is not important to the evolution of the lobe, and the role of the jet is simply to supply energy to the synchrotron-emitting lobe. 
		The subsequent expansion of the lobes is determined by solving a set of fluid conservation equations; typically, the lobes undergo an initial momentum-dominated supersonic phase, followed by an adiabatic-expansion-driven coasting phase, and ultimately rising buoyantly in the later phases of evolution.}
	
	A large number of analytical and numerical models describing the evolution of active galactic nucleus (AGN) jets and lobes have been published since the first models were introduced over five decades ago. In this review, we summarise the different classes of lobed radio galaxy models, and provide a common framework to facilitate comparison both between the model classes and to more detailed hydrodynamic simulations. The dynamics of jetted Fanaroff and Riley~{\cite{Fanaroff+1974}} Type-I (FR-I) sources are not considered in this work; we refer the interested reader to the classical work of \citet{Bicknell+1995}.
	
	The review is structured as follows.
	Section \ref{sec:The First Models} presents early models by \citet{Rees+1971} {and} \citet{Scheuer+1974} (Model A), in which the forward thrust of uncollimated jets is balanced by the ram pressure from the ambient medium. We extend the formalism of \citet{Scheuer+1974} to non-uniform environments, enabling direct comparison with more sophisticated modern analytical models for the first time.
	While these early models capture the fundamental aspects of jet termination and lobe formation, they neglect jet collimation by sideways ram pressure from the lobe (or ambient medium). In Section \ref{sec:Lobe Expansion Models}, we describe the analytical models proposed by \citet{Falle+1991} and \citet{KA+1997}, which link jet collimation to subsequent lobe expansion.
	The past decade has seen the advent of environment-sensitive radio galaxy models beyond the self-similar solutions of \citet{KA+1997} and related models ({e.g.,} \citep{Blundell+2000, Manolakou+2002}). These models capture the evolution of lobe morphology in realistic environments \citep{Turner+2015, Hardcastle+2018} as well as the transition between jet- and lobe-driven expansion \citep{Hardcastle+2018, Turner+2023}. We describe these models in Section \ref{sec:Numerical Models}.
	In Section \ref{sec:Discussion}, we compare the consistency of predicted radio source dynamics between the main model classes, benchmark their evolutionary tracks against hydrodynamic simulations, and discuss their ability to generate synthetic AGN populations for parameter inversions. We conclude and suggest improvements to implement in the next generation of analytical models in Section \ref{sec:Conclusions}.

	\section{Early Jet--Lobe Models}
	\label{sec:The First Models}
	

	\citet{Rees+1971} proposed a model in which conical jets emanating from the central engine of active galactic nuclei (AGNs) are pressure-balanced by the ram pressure of the ambient medium.
	In this model, the jets are a beam of low-frequency electromagnetic waves (LFEMW), the quantum field equivalent of a pair-plasma. The radiation pressure of this beam upon absorption by the ambient medium is $p_{rad} = Q/(\Omega R^2 c)$ for jet kinetic power $Q$ and beam cross-sectional area $\Omega R^2$ at radius $R$ from the active nucleus. The pressure is increased if the interaction between the beam and ambient medium results in pair production, leading to a reaction pressure up to double that of the radiation pressure; the exact factor depends on the angle of reflected particles. The pressure contribution from the jet is, therefore, expressed as:
	\begin{equation}
		p_{jet} = \frac{\kappa_1 Q}{\Omega R^2 c} ,
		\label{scheuerjet}
	\end{equation}
	where $1 \leqslant \kappa_1 < 2$ is a dimensionless constant describing both the fraction of the beam power that interacts with the ambient medium as radiation or particles, and the angle of reflection of those particles.
	
\begin{figure}[H]
		
		\includegraphics[width=0.7\textwidth,trim={0 0 0 0},clip]{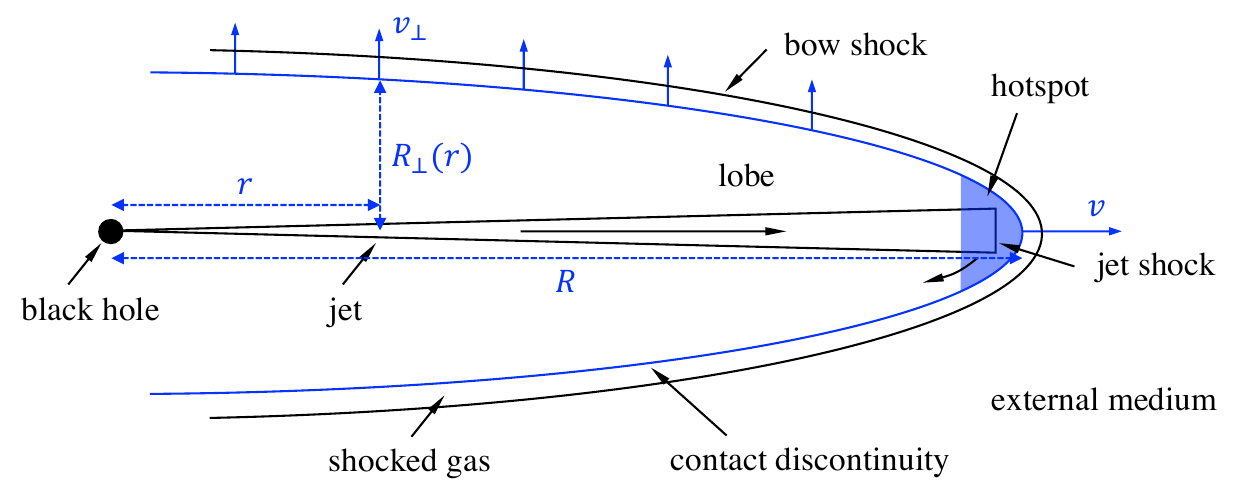}
		
		\caption{Schematic of the dynamical model for the \citet{Scheuer+1974} Model A. We show a thin shocked gas shell between the contact discontinuity and bow shock as in Figure \ref{fig:scheuer_dynamics} of the original paper; however, the shocked gas is not explicitly considered in their model.}
		\label{fig:scheuer_dynamics}
	\end{figure}

	\textls[-25]{The forward jet thrust is balanced by ram pressure from the ambient medium\mbox{ $p_{ram} = \rho v^2$}, where $\rho$ is the gas density of the assumed constant density ambient medium and $v = dR/dt$ is the advance speed of the jet head (see Figure \ref{fig:scheuer_dynamics}). \citet{Scheuer+1974} evaluated the resulting first-order differential equation in $R$ assuming a constant jet half-opening angle $\theta_j$ (and thus, solid angle $\Omega$) in their Model A. The jet length is related to source age $t$, and jet and environment parameters~as:}
	\begin{equation}
		R(t) = \left(\frac{\kappa_1 Q}{\Omega \rho c}\right)^{1/4} \left(2t\right)^{1/2} .
	\end{equation}
	{The} 
	above approach assumes a constant density environment. However, the ambient medium on scales exceeding several kiloparsecs---typical of extended radio sources---is well represented by a symmetric power-law density profile of the form $\rho = k r^{-\beta}$, where the density parameter $k \equiv \rho_0 r_0^\beta$ is a constant (e.g., \citep{Falle+1991,Turner+2015}). The ram pressure applied by the ambient medium onto the expanding jet consequently weakens with distance from the central nucleus. In this review, we extend the \citet{Scheuer+1974} Model A for the \textit{more} general case of a power-law density profile, yielding:
	\begin{equation}
		R(t) = \left(\frac{\kappa_1 Q}{\Omega k c}\right)^{1/(4-\beta)} \left(\frac{(4-\beta)t}{2} \right)^{2/(4-\beta)} ,
		\label{scheuer}
	\end{equation}
	which converges to Scheuer's {\cite{Scheuer+1974}} original constant density form when $\beta = 0$, noting that $\rho = k$ in this limiting case. We use this more complete version of the model in the remainder of this work.
	\vspace{-6pt}
	
	The synchrotron-emitting lobes inflated by the jets are typically assumed to have ellipsoidal morphology, with the ratio of major (aligned with the jet) to minor axes defined by the axis ratio $A = R/R_\perp$; we note that this differs (by a factor of 2) from the \textit{axial} ratio $R_T = A/2$ of \citet{KA+1997}.
	\citet{Scheuer+1974} derive the volume of the ellipsoidal radio lobe associated with their LFEMW jets by considering the work done in inflating the cavity. The total energy in the cavity, $U$, increases over the time interval $\delta t$ due to the input kinetic power $Q$ as:
	\begin{equation}
		\delta U = Q \delta t - p \delta V ,
		\label{energyeq}
	\end{equation}
	where $\delta V$ is the differential increase in volume, and the lobe pressure is given as follows (see, e.g.,~\citep{KA+1997}, specifically their Equation (15)):
	\begin{equation}
		p = \frac{U(\Gamma_c - 1)(q + 1)}{V} ,
		\label{udensity}
	\end{equation}
	%
	%
	%
	where $\Gamma_c$ is the polytropic index (or adiabatic index for an adiabatic equation of state; EoS) of the lobe plasma, and $q \ll 1$ is the ratio of energy in the magnetic field to that in the particles. This equation assumes the energy density (and thus, pressure) is approximately uniform throughout the lobe, a reasonable assumption given the high ($\sim$$0.1c$) sound speeds in the lobes.
	
	Equation \eqref{energyeq} is a first-order differential equation describing the evolution of the total energy of the cavity. In Appendix \ref{app:Lobe pressure}, we solve this differential equation assuming the cavity volume expands with increasing jet length as $V(R) = \kappa_2 R^\alpha$; here $\alpha, \kappa_2 > 0$ are constants. This yields an expression for the lobe pressure in terms of the source age:
	\begin{equation}
		p(t) = \frac{Q (\Gamma_c - 1)(q + 1)}{\kappa_2 [\alpha(\Gamma_c - 1)(q + 1) + (4 - \beta)/2] } \left(\frac{\Omega k c}{\kappa_1 Q}\right)^{\alpha/(4-\beta)} \left(\frac{(4-\beta)t}{2} \right)^{(4 - \beta - 2\alpha)/(4-\beta)} ,
		\label{pressse}
	\end{equation}
	where the constants $\alpha$ and $\kappa_2$ are evaluated below by considering the lobe volume evolution.
	
	A major limitation of the \citet{Scheuer+1974} model concerns the sideways expansion of the lobe, which is assumed to occur at the same velocity at every point on the lobe surface at any given time $t$. This expansion rate is derived by equating the lobe pressure to the ram pressure presented by the ambient medium as the lobe widens, i.e., $\rho v_\perp^2 = p(t)$, where the ambient gas density is reasonably approximated as $\rho \sim kR^{-\beta}$.
	This sidewards expansion can only commence at locations already reached by the jet material. The half-width of the lobe at some location $r$ along the jet axis is, thus, given by:
	\begin{equation}
		R_\perp(r) = \int_{t(r)}^{t(R)} v_\perp(t^*) dt^* ,
	\end{equation}
	where $t(r)$ is the time when the jet head reached the location $r$ along the jet axis and $t(R)$ is the source age when the jet has its present length $R$. This integral is evaluated in Appendix~\ref{app:Lobe volume}.
	
	The lobe volume at the source age $t \equiv t(R)$ is found by integrating over all locations $r$ along the jet axis:
	\begin{equation}
		\begin{split}
			V(R) \equiv \kappa_2 R^\alpha &= \int_0^R \pi R_\perp(r)^2 dr \\
			&\propto R^{(14 - \beta - 2\alpha)/2} ,
		\end{split}
		\label{volsh}
	\end{equation}
	{Dimensional} 
	analysis shows that the only possible solution is for $\alpha = (14 - \beta)/4$. The lobe expansion is, therefore, {\it not} self-similar (which would require $V \propto R^3$, and hence, $\alpha=3$) unless $\beta=2$, i.e., in a rapidly declining density profile representative of the outer regions of groups or clusters. For a uniform medium, as considered by \citet{Scheuer+1974}, the exponent converges to $\alpha = \tfrac{7}{2}$.
	
	The constant of proportionality, $\kappa_2$, is similarly found by comparing terms not involving $R$, yielding:
	\begin{equation}
		\kappa_2 = \frac{16\pi^{1/2} (\Omega c)^{3/4} k^{1/4}}{[(14 - \beta)(18 - \beta)]^{1/2} \kappa_1^{3/4} Q^{1/4}}  \left[\frac{(\Gamma_c - 1)(q + 1)}{(14 - \beta)(\Gamma_c - 1)(q + 1) + 2(4 - \beta)} \right]^{1/2} ,
	\end{equation}
	which converges to the expression found by (\citet{Scheuer+1974}, specifically their Equation (10)) in the limit of a uniform ambient medium and assuming $\Gamma_c = \tfrac{4}{3}$. 
	
	This simple model neglects the sidewards ram pressure of the ambient medium (or lobe at later times) acting on the jet, which will lead to reconfinement shocks and ultimately the collimation of the jet. \citet{Scheuer+1974} proposed a second model (their Model B) in which the jet is smoothly compressed into a collimated beam by the ambient medium; however, this assumption leads to unphysically narrow jets and consequently significantly faster jet-head advance speeds, which scale with jet cross-section $y$ as $R(t) \propto y^{-4/9} t^{7/9}$. {\mbox{\citet{Scheuer+1982}} {subsequently} 
		proposed that, in some sources, the jet may precess on a timescale which is short compared to the evolutionary timescale of the lobe. The time-averaged momentum flux of the jet is effectively spread over a larger cross-sectional area (equivalent to a larger jet opening angle), resulting in a slower growth rate along the jet axis (e.g., \citep{Begelman+1989})}. We return to this point in Section~\ref{sec:Discussion}, when we compare the predictions of different models.

	\section{Lobe Expansion Models}
	\label{sec:Lobe Expansion Models}
	
	The self-similar expansion model for the growth of quasar winds by \mbox{\citet{Dyson+1980}} spurred a new generation of analytical models based on the adiabatic expansion of the lobe bubble along a power-law ambient gas density profile. In particular, \citet{Falle+1991} related the geometry and internal pressure of the expanding lobe to the dynamics of the jet (\mbox{Section \ref{sec:Jet collimation}}), enabling the \citet{Dyson+1980} model to be modified to consider the evolution of radio sources (Section \ref{sec:Lobe adiabatic expansion}). The \citet{Falle+1991} model forms the basis for several models in the literature, including those by \citet{KA+1997}, \citet{Blundell+2000}, and \citet{Manolakou+2002}.
	
	\subsection{Jet Collimation}
	\label{sec:Jet collimation}
	
	\citet{Falle+1991} revisited the dynamical modelling of jet collimation in their work in 1999 {\cite{Falle+1991}}, considering an initially conical jet that reflects a strong shock of 
	the surrounding medium upon reaching lateral pressure equilibrium. This reconfinement shock bounces between each side of the jet cavity preventing any further decay of the lateral thrust against the surrounding medium; this leads to a constant width, collimated jet with repeated cross-shaped structures of enhanced pressure and synchrotron emissivity.
	\citet{Falle+1991} assumed that lobe formation occurs prior to jet collimation, and thus, that it is the lobe pressure which opposes the sidewards component of the jet thrust, not the ambient medium. \citet{Alexander+2006} showed that jets may in fact be collimated by the ambient medium prior to the formation of lobes if $\sqrt{\Gamma_x}\sin\theta_j M_x < 1$, where $\Gamma_x$ is the polytropic index of the ambient medium, $\theta_j$ is the half-opening angle of the conical jet, and $M_x$ is the Mach number of the jet-head advance with respect to the sound speed of the ambient medium. Jet-head advance speeds are initially relativistic (e.g., VLBI observations of \citep{Britzen+2008}), and thus, the external Mach number of the jet, is expected to be of an order of several hundred; only for very small opening angles of $<$1~degree would the jet be expected to be collimated by the ambient medium rather than the lobe.
	
	The location of the initial reconfinement shock, $z$, is found by applying the Rankine--Hugoniot jump conditions for a plane-parallel shock to the lateral component of the flow that travels along the jet edge. The pressure of the lobe plasma is related by these conditions to the lateral component of the jet thrust as:
	\begin{equation}
		p(t) = \frac{2}{\Gamma_j + 1} \rho_j(z_1) v_j^2 \sin^2\theta_j,
		\label{jetpress}
	\end{equation}
	where the jet plasma has a bulk velocity $v_j$, polytropic index $\Gamma_j$, and density $\rho_j(z_1)$ at the critical radius $z_1$ at which the jet begins to collimate (i.e., location where the ram pressure first matches the lobe pressure). The density of the jet plasma will remain constant after this point until it reaches the jet head. We can, thus, derive an expression for the pressure acting on the contact discontinuity between the jet head and surrounding shocked gas using the shock jump conditions. That is:
	\begin{equation}
		\begin{split}
			p_h(t) &= \frac{2}{\Gamma_j + 1} \rho_j(z_1) v_j^2 \\
			&= \frac{p(t)}{\sin^2 \theta_j},
		\end{split}
		\label{pressratio}
	\end{equation}
	where the second equality is obtained upon substitution of Equation \eqref{jetpress}. This expression has the same form as found by \citet{KA+1997}  (specifically their Equation (36)), and yields comparable pressure ratios to the numerically informed value obtained by \citet{Komissarov+1998}.
	
	The lobe and jet-head region are surrounded by a shell of swept-up ambient medium that has been overrun by the bow shock generated by the expanding jet. This shocked gas is in approximate pressure equilibrium with the proximate lobe/jet-head plasma, but has significantly higher density, and thus, lower temperature (e.g., \citep{Hardcastle+2013, Yates+2021}), i.e., $p_s(t) \sim p_h(t)$, where $p_s(t)$ is the pressure just inside the bow shock along the jet axis. Together with conservation equations, these relationships between the conditions in the lobe cavity, the bow shock, and the ambient medium are sufficient for describing the evolution of the expanding radio source.

	\subsection{Lobe Adiabatic Expansion}
	\label{sec:Lobe adiabatic expansion}

	\citet{Falle+1991} presented their model in terms of the volume and pressure of the lobe; however, the surrounding shocked gas shell also receives a non-negligible fraction of the input energy from the central nucleus. Therefore, we express their equations in a more complete form considering both the lobe and shocked gas shell consistent with later work by \mbox{\citet{Hardcastle+2018}}, \citet{Turner+2020a}, and \citet{Turner+2023}. These models assume that it is primarily the thermal pressure of the lobe plasma that drives the source expansion; this pressure is uniform throughout the lobes due to high internal sound speed. The first law of thermodynamics relates the jet kinetic energy input to the thermal pressure $p$ in the lobe and shell, and shocked gas volume $V_s$ (see Equation (6) of \citep{Alexander+2006}):
	\begin{equation}
		V_s \frac{dp}{dt} + \Gamma_c p \frac{dV_s}{dt} = (\Gamma_c - 1)Q ,
		\label{gov}
	\end{equation}
	where $\Gamma_c$ is the adiabatic index of the shocked gas and lobe plasma ($\tfrac{5}{3}$ for a non-relativistic fluid), and $Q$ is the power injected into the shocked shell by the jet. This equation can be rewritten in terms of the pressure at the interface of the shocked shell and ambient medium (along the jet axis) using Equation \eqref{pressratio}. That is:
	\begin{equation}
		V_s \frac{dp_s}{dt} + \Gamma_c p_s \frac{dV_s}{dt} = \frac{(\Gamma_c - 1)Q}{\sin^2\theta_j} .
		\label{gov2}
	\end{equation}

\begin{figure}[H]
		
		\includegraphics[width=0.8\textwidth,trim={0 0 0 0},clip]{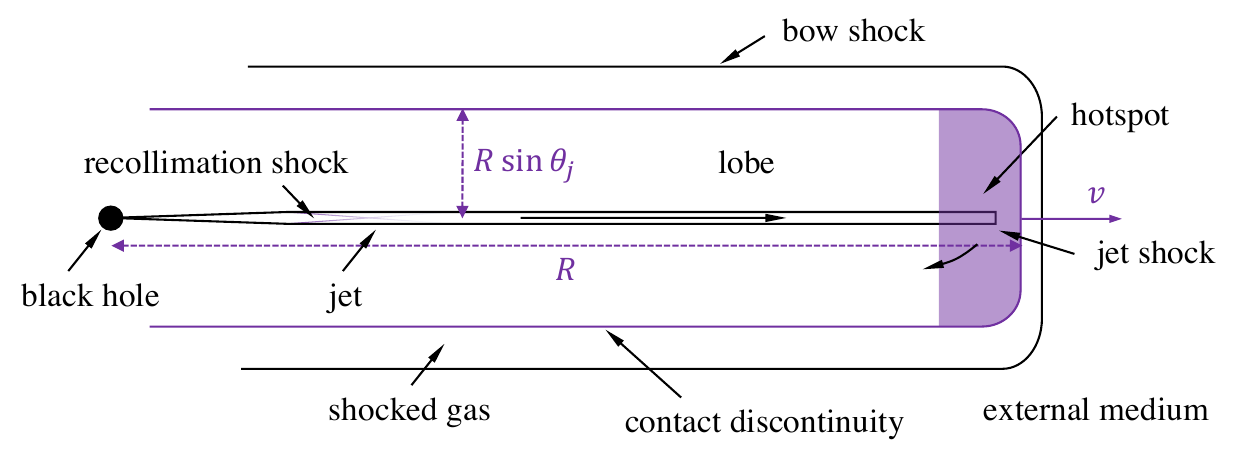}
		
		\caption{Schematic of the dynamical model proposed by \citet{Falle+1991}, and subsequently refined by others, including \citet{KA+1997} and \citet{Alexander+2006}. The bow shock is assumed to expand in a self-similar manner (i.e., constant scaling to the lobe) but the energy associated with the shocked gas is not explicitly considered.}
		\label{fig:falle_dynamics}
	\end{figure}

	\citet{Falle+1991} showed that the shocked shell expands in a self-similar manner, leading to a constant scaling between the volume and cube of the jet length. That is:
	\begin{equation}
		V_s(R_s) = \kappa_3 R_s^3 .
	\end{equation}
	where $R_s$ is the radius of the shocked gas shell along the jet axis and $\kappa_3$ is a constant of proportionality. \citet{KA+1997}, and some subsequent authors (e.g., \citep{Alexander+2000}), modelled the lobe as cylindrical (see Figure \ref{fig:falle_dynamics}), with the major axis of the shocked gas shell being a factor of $A_s = 1/\sin \theta_j$ longer than the minor axis. The volume of the shocked shell is then:
	\begin{equation}
		V_s(R_s) = \pi \sin^2 \theta_j R_s^3 \ \ \ \text{(cylinder)} .
		\label{shockvol}
	\end{equation}
	{In} this approach, expansion in the jet direction is driven by the ram pressure acting on the jet-head region, $p_h(t)$, whilst sideways expansion is driven by shocked shell thermal pressure, $p(t)$. The more realistic assumption of an ellipsoid shocked shell adds a factor of $\tfrac{2}{3}$ to the above equation; however, the sidewards ram pressure then becomes a strong function of distance along the jet axis. The resulting \textit{average} lobe and shocked shell pressures must be calculated numerically (see Section \ref{sec:RAiSE angles}) rather than using the simple relation derived in Section \ref{sec:Jet collimation}.
	\vspace{-6pt}
	
	The pressure of the shocked shell along the jet axis, $p_s(t) \sim p_h(t)$, is related to the density of the ambient medium, $\rho_x = kR_s^{-\beta}$, using the Rankine--Hugoniot shock jump conditions:
	\begin{equation}
		p_s(t) = \frac{2}{\Gamma_x + 1} kR_s^{-\beta} \left(\frac{dR_s}{dt}\right)^2,
		\label{shockpress}
	\end{equation}
	where $\Gamma_x$ is the adiabatic index of the ambient medium surrounding the shocked gas shell. 
	
	Substituting Equations \eqref{shockvol} and \eqref{shockpress} for the shocked shell volume and pressure along the jet axis into Equation \eqref{gov2} yields a second-order non-linear differential equation for the shell radius:
	\begin{equation}
		2R_s^{3-\beta} \frac{dR_s}{dt} \frac{d^2R_s}{dt^2} + (3\Gamma_c - \beta) R_s^{2-\beta} \left(\frac{dR_s}{dt}\right)^3  = \frac{(\Gamma_c - 1)(\Gamma_x + 1)Q}{2k \pi \sin^2\theta_j} .
		\label{diff}
	\end{equation}
	{This} equation can be solved by trialling another power-law solution, with the exponent again constrained by dimensional analysis. This yields:
	\begin{equation}
		R_s(t) = \left[\frac{(\Gamma_c - 1)(\Gamma_x + 1)(5 - \beta)^3 Q}{18 (9\Gamma_c - 4 - \beta) k \pi \sin^2\theta_j} \right]^{1/(5 - \beta)} t^{3/(5 - \beta)} .
		\label{shockradius}
	\end{equation}
	{\citet{KA+1997}} included an additional correction in the denominator of their equivalent expression due to the energy associated with the higher pressure jet-head region; the $(9\Gamma_c - 4 - \beta)$ term in Equation \eqref{shockradius} becomes $(9[\Gamma_c + (\Gamma_c - 1)/\sin^2\theta_j] - 4 - \beta)$. We note that this correction assumes that the differential increase in volume of the lobe and hotspot are equal as the source grows; this assumption is not particularly realistic for an ellipsoidal lobe geometry.

	\section{Semi-Analytic Models}
	\label{sec:Numerical Models}
	
	Improved computation has recently enabled a new generation of analytical models with added complexity. These models typically solve systems of differential equations which lack an analytic solution to describe the evolutionary history of the radio source. Below, we summarise the main developments, including atmospheres beyond power-law density profiles (Section \ref{sec:RAiSE angles}), considering both the ram and thermal pressure contributions to the expansion along the jet axis (Section \ref{sec:Hertfordshire model}), and modelling the relativistic jet in a distinct expansion phase prior to the onset of lobe formation (Section \ref{sec:RAiSE HD}).
	
	\subsection{RAiSE (Version 2015)}
	\label{sec:RAiSE angles}
	
	\citet{Turner+2015} developed a semi-analytic model, \emph{Radio AGN in Semi-analytic Environments} (RAiSE), based on the theory of the \citet{Falle+1991} class of models (Section \ref{sec:Lobe Expansion Models}). These authors extended existing analytic approaches by considering piece-wise solutions to the governing differential equations in two dimensions. The RAiSE model included three key improvements over the earlier models: (1) ambient medium consistent with X-ray observations of clusters and semi-analytic galaxy formation models, (2) angular dependence of expansion velocity across the ellipsoidal contact surface, and (3) modelling of the morphological transition from supersonic to subsonic lobe expansion by using complete differential equations rather than limiting cases, which yield analytic expressions.

\begin{figure}[H]
		
		\includegraphics[width=0.8\textwidth,trim={0 0 0 0},clip]{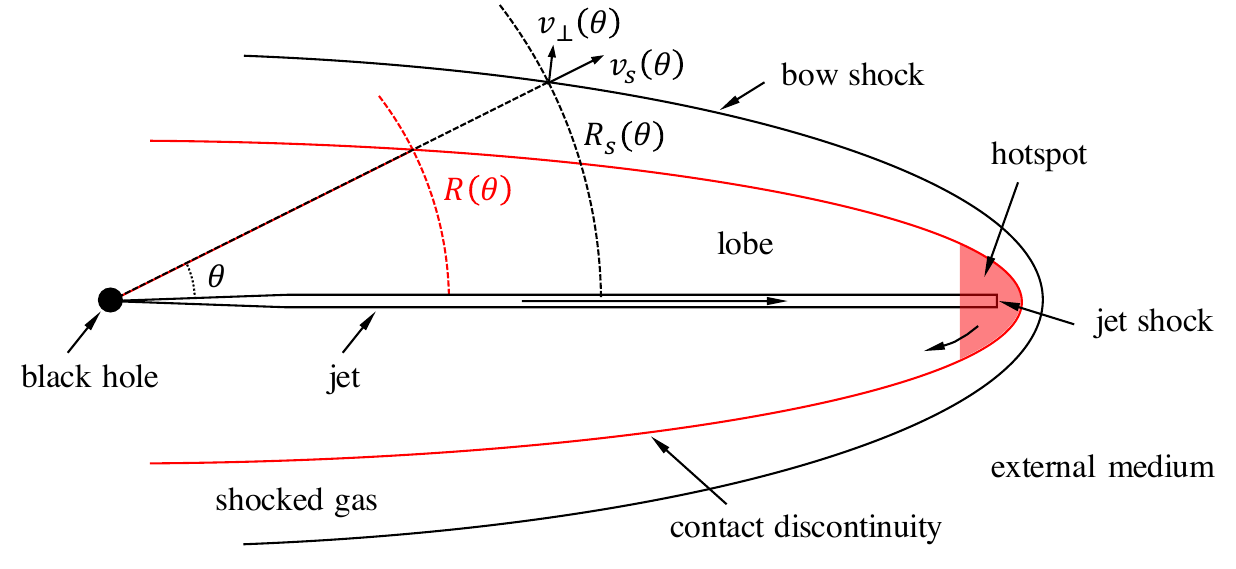}
		
		\caption{Schematic of the \citet{Turner+2015} dynamical model for the lobe and shocked shell. This framework is also used by \citet{Turner+2023} for both their jet- and lobe-dominated expansion phases, albeit the lobe (shown in red) only forms once a critical length scale is reached. {{Taken} 
				from Figure 1 of \citet{Turner+2023}.}}
		\label{fig:dynamics}
	\end{figure}

	The lobe and shocked shell in their model are constructed from an ensemble of small angular volume elements in assumed pressure equilibrium. Each element of fixed angular width $d\theta$ is assumed to receive a constant fraction of the jet power as the cavity expands. This assumption yields self-similar expansion at early times when the shocked shell is expanding in the strong-shock supersonic limit, as in the earlier models of\break \citet{KA+1997}. The volume of each small angular element of the shocked shell, $[\theta - \delta\theta/2, \theta + \delta\theta/2)$, is given by:
	\begin{equation}
		\delta V_s(\theta) = \frac{2\pi R_s^3(\theta)}{3} \sin\theta \delta\theta,
		\label{vol2}
	\end{equation}
	where $\theta$ is the angle between some location on the surface of the shocked shell and the jet axis and $R_s(\theta)$ is the radius of the \textit{initially} ellipsoidal shell at that location (see Figure \ref{fig:dynamics}). Importantly, the shocked gas shell does not expand self-similarly as the steepness of the ambient gas density profile encountered by the non-spherical shell will in general differ across its surface, leading to different growth rates; this prediction is consistent with the higher axis ratios observed in the largest radio sources \citet{Mullin+2008}.
	\vspace{-3pt}

	The initial radius of each volume element is related to that along the jet axis by a geometric factor as $R_s(\theta, t \rightarrow 0) = \eta_s(\theta) R_s(\theta=0, t \rightarrow 0)$, where $\theta = 0$ is aligned along the jet axis and $\eta_s(\theta)$ is defined as:
	\begin{equation}
		\eta_s(\theta) = \frac{1}{\sqrt{(\sin^2\theta/\sin^2\theta_j) + \cos^2\theta}} ,
		\label{etas}
	\end{equation}
	where we have assumed the same relationship between the jet half-opening angle, $\theta_j$, and axis ratio of the shocked shell, $A_s$, as discussed in Section \ref{sec:Lobe Expansion Models}.
	
	Following \citet{Turner+2015}, and later work by \citet{Turner+2020a} and \citet{Turner+2023}, the adiabatic expansion of each angular volume element is related to the pressure imparted on that element at the surface, $p_s(\theta)$, its volume $\delta V_s(\theta)$, and the fraction of the input jet power associated with that element, $Q \delta\lambda(\theta)$. The function $\delta\lambda(\theta)$ is defined in Equation 20 of \citet{Turner+2023}. The first law of thermodynamics in Equation \eqref{gov2} gives:
	\begin{equation}
		\frac{dp_s(\theta)}{dt} \delta V_s(\theta) + \Gamma_c p_s(\theta) \frac{d[\delta V_s(\theta)]}{dt} = (\Gamma_c - 1)Q \delta\lambda(\theta) .
		\label{gov3}
	\end{equation}
	{Away} from the contact surface, the pressure in the lobe is calculated as the spatial average of the surface pressures $p_s(\theta)$. 
	
	\citet{Turner+2015} used a similar expression for the pressure at the contact surface to Equation \eqref{shockpress}, but additionally considered: (1) the orientation of the expanding surface as it impacts the ambient medium and (2) terms describing evolution in the transonic and subsonic expansion regimes. That is, for expansion in the supersonic and transonic phases, we have:
	\begin{equation}
		p_s(\theta) = \frac{2}{\Gamma_x + 1} kR_s^{-\beta}(\theta) \left(\frac{\zeta_s(\theta)}{\eta_s(\theta)} \frac{dR_s(\theta)}{dt}\right)^2 - \frac{\Gamma_{x} - 1}{\Gamma_{x} + 1} (k l)R_s^{-\beta}(\theta) ,
		\label{shockjump2}
	\end{equation}
	where $\zeta_s(\theta)$ is a further geometric factor defined in Equation (13) of \citet{Turner+2023}. The radial temperature profile of the ambient medium is defined by \citet{Turner+2015} as $T = (\bar{m}/k_b) l r^{-\varsigma}$ for Boltzmann constant $k_B$, and average particle mass $\bar{m} \sim 0.6 m_p$ (where $m_p$ is the proton mass). Below, we present their results assuming an isothermal medium for consistency with other authors, i.e., adopting $\varsigma = 0$.
	The pressure in the subsonic regime is equal to the ambient pressure (see Equation (4) of \citep{Turner+2015}).
	
	\textls[5]{The second-order non-linear differential equation that results from substituting \mbox{Equations \eqref{vol2} and \eqref{shockjump2}} into Equation \eqref{gov3} cannot in general be solved to yield an analytic solution. \citet{Turner+2015}, instead, rewrote the resulting equation as a system of two coupled first-order ordinary differential equations. These differential equations describe the velocity and acceleration at the contact surface of a given volume element $\delta V_s(\theta)$:}
	\begin{equation}
		\begin{split}
			\dot{R}_s(\theta) &= v_s \\
			\dot{v}_s(\theta) &= \frac{3 (\Gamma_{x} + 1)(\Gamma_{c} - 1) Q R_s^{\beta - 3} \delta\lambda}{8 \pi v_s (\zeta_s/\eta_s)^2 k \sin\theta \delta\theta} + \frac{(\beta - 3\Gamma_{c}) v_s^2}{2 R_s} \\
			&\quad\quad + \frac{(\Gamma_{x} - 1) (3 \Gamma_{c} - \beta) l}{4 R_s (\zeta_s/\eta_s)^2} ,
		\end{split}
		\label{supersonic system}
	\end{equation}
	where $v_s$, $R_s$, $\delta\lambda$, $\zeta_s$, and $\eta_s$ are explicit functions of $\theta$, whilst the properties of the ambient medium, $k$, $l$ and $\beta$, are implicit functions of $\theta$ as different sections of the contact surface reach a given distance into the spherically symmetric environment at different times. \citet{Turner+2015} used a standard fourth-order Runge--Kutta method to solve this system of equations, providing the analytic solution for the strong-shock limit as an initial condition.
	
	\subsection{Hardcastle Model}
	\label{sec:Hertfordshire model}

\begin{figure}[H]
		
		\includegraphics[width=0.8\textwidth,trim={0 0 0 0},clip]{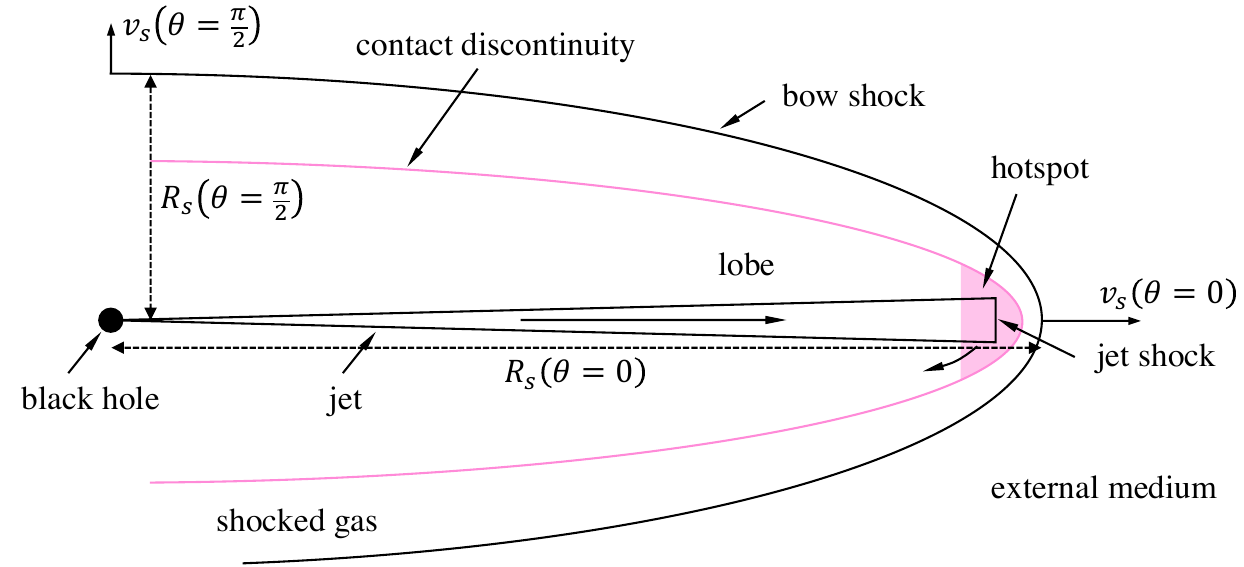}
		
		\caption{Schematic of the dynamical model proposed by \citet{Hardcastle+2018}. We depict a conical jet, noting that in this model, the cross-sectional area at the jet head is related to the lobe volume/radius by a constant scaling factor $\kappa_1$, and hence, the dynamics of the jet are not critical to model behaviour.}
		\label{fig:hardcastle_dynamics}
	\end{figure}

	\citet{Hardcastle+2018} improved on the lobe-dominated expansion model of \citet{Turner+2015} by explicitly considering the momentum flux of the jet plasma. This model considers expansion for two angles across the surface of the shocked shell, $\theta = 0$ and $\tfrac{\pi}{2}$ (see Figure \ref{fig:hardcastle_dynamics}). The choice of these two angles is sufficient to model changes to the axis ratio of the lobe and shocked shell in lobed FR-IIs, with the larger number of angles considered by \citet{Turner+2015} only important in the transonic and subsonic expansion phases when the lobe deforms from its initial ellipsoidal shape.
	
	\vspace{-6pt}
	
	The shocked shell pressures are derived from the ram pressure component along the jet axis (Equation \eqref{scheuerjet}), and a component due to the internal energy of the relativistic lobe plasma acting along both axes (Equation \eqref{udensity}, with $U = Qt$ and $q \ll 1$). The lobe and shocked gas are assumed to be in pressure equilibrium, as discussed in Section \ref{sec:Lobe Expansion Models}. The shocked shell pressures along the major and minor axes are then given by:
	\begin{equation}
		\begin{split}
			p_s(\theta=0) &= \frac{\varepsilon Q R_s(\theta=0)}{2cV} + \frac{(\Gamma_c - 1)\xi Qt}{V} \\
			p_s(\theta=\tfrac{\pi}{2}) &= \frac{(\Gamma_c - 1)\xi Qt}{V} .
		\end{split}
		\label{hardcastle pressure}
	\end{equation}
	where $V/R_s(\theta=0)$ is the cross-sectional area of the jet-head region for a cylindrical lobe of volume $V$, $\varepsilon \sim 4$ here acts as a geometric correction factor reflecting more realistic lobe shapes, and $\xi \sim \tfrac{1}{2}$ is the fraction of the input jet kinetic power found in hydrodynamic simulations to be stored as internal energy of the relativistic lobe plasma; the remainder is stored as thermal and kinetic energy in the shocked gas shell.
	
	\citet{Hardcastle+2018} related the volume of the lobe, $V$, to that of the shocked shell, $V_s$ (including the interior lobe), by considering the ratio of total internal energies:
	\begin{equation}
		\frac{V}{V_s} = \frac{(\Gamma_c - 1)\xi Qt}{[\xi \Gamma_c + (1 - \xi)\Gamma_s - 1]Qt + f(N, T, v_s)} ,
		\label{Vh}
	\end{equation}
	where $f(N, T, v_s)$ is a function describing the internal energy of the ambient medium swept up by the bow shock. This function depends on the total number of swept-up particles, $N$, their temperature, $T$, and their bulk velocity due to the expansion of the shocked shell, $v_s = dR_s/dt$ (\citep{Hardcastle+2018}, specifically their Equations (5) and (7)). For young sources, when the thermal energy of these particles is lower than the energy supplied by the jet to the forming shocked gas shell, the ratio of lobe to shocked shell volumes tends to $\xi$ for $\Gamma_c = \Gamma_s = \tfrac{5}{3}$, or $\xi/(2-\xi)$ if the lobe is assumed to have a relativistic plasma ($\Gamma_c = \tfrac{4}{3}$). The volume of the shocked gas shell is of course also directly related to the lengths of the major and minor axes; for an ellipsoidal geometry this gives:
	\begin{equation}
		V_s = \frac{2\pi R_s(\theta = 0)R_s^2(\theta = \tfrac{\pi}{2})}{3} .
		\label{Vsh}
	\end{equation}

	{We can, therefore, express the pressure along the major and minor axes of the shocked shell (Equation \eqref{hardcastle pressure}) as a function of {{both}} 
		axis lengths upon substitution of \mbox{Equations \eqref{Vh} and \eqref{Vsh}}.}
	
	\citet{Hardcastle+2018} rewrote the Rankine--Hugoniot shock jump conditions (Equation \eqref{shockjump2}) in terms of the velocities along the major and minor axes of the lobe. However, unlike the previously discussed models, \citet{Hardcastle+2018} expressed the jump conditions in terms of the sound speed, $c_s$, and adiabatic index, $\Gamma_s$, of the shocked gas shell surrounding the lobe. The expansion rates along the major and minor axes are then given by:
	\begin{equation}
		\begin{split}
			\frac{dR_s(\theta = 0)}{dt} &= c_s \sqrt{\frac{(\Gamma_s + 1)[p_s(\theta = 0)/p_x(R_s(\theta = 0))] - (\Gamma_s - 1)}{2\Gamma_s}} \\
			\frac{dR_s(\theta = \tfrac{\pi}{2})}{dt} &= c_s \sqrt{\frac{(\Gamma_s + 1)[p_s(\theta = \tfrac{\pi}{2})/p_x(R_s(\theta = \tfrac{\pi}{2}))] - (\Gamma_s - 1)}{2\Gamma_s}} ,
		\end{split}
		\label{hardcastle system}
	\end{equation}
	where $p_x(r)$ is the spherically symmetric ambient gas pressure profile. This coupled system of non-linear ordinary differential equations can be solved using a standard fourth-order Runge--Kutta method, with initial conditions of $R_s(\theta = 0) = ct_0$ and $R_s(\theta = \tfrac{\pi}{2}) = ct_0$ for some small time $t_0$.

	\subsection{RAiSE (Version 2023)}
	\label{sec:RAiSE HD}
	
	The RAiSE \citep{Turner+2015} model discussed in Section \ref{sec:RAiSE angles} was first extended to make predictions for the spatial distribution of emission at radio \citep{Turner+2018a} and X-ray wavelengths \citep{Turner+2020a}, and was most recently used to incorporate important changes to jet and lobe dynamics. The \citet{Turner+2023} model includes: (1) a relativistic jet expansion phase modelled prior to the formation of a lobe, (2) formation of lobes within a surrounding bow shock, and (3) a separation of the ram and thermal components of the jet and lobe pressure.
	
	The \citet{Turner+2023} model is implemented using the same computational framework as the original RAiSE model discussed in Section \ref{sec:RAiSE angles}, specifically, using coupled differential equations which are solved for small angular volume elements of the lobe and shocked shell. Below, we present a concise derivation of their solution for the expansion of the relativistic jet (Section \ref{sec:Relativistic hydrodynamic equations}), summarise their methodology to model the subsequent lobe formation/inflation (Section \ref{sec:Lobe formation}), and finally present their method to separate the ram and thermal components of the jet and lobe pressure (Section \ref{sec:Thermal pressure}). 
	
	\subsubsection{Relativistic Jet Expansion}
	\label{sec:Relativistic hydrodynamic equations}
	
	The relativistic hydrodynamic conservation equations relate the properties of fluids upstream and downstream of a shock discontinuity via the stress-energy tensor. The conservation equations for a relativistic fluid are expressed in terms of comoving quantities including gas density $\rho$, gas pressure $p$, dimensionless specific enthalpy $h$ (i.e., enthalpy divided by $c^2$), and the non-zero spatial component of the four-velocity $u = \gamma v/c$ (hereafter shortened to four-velocity) relative to the shock front. 
	The conservation equations for a relativistic fluid are {({e.g., } 
		\citep{Turner+2023}, and references therein)}{:}
	\begin{subequations}
		\begin{gather}
			\rho \gamma vs. = \rho_1 \gamma_1 v_1 \ \ \ \text{(continuity)} \label{continuity}\\
			\rho h \gamma^2 v^2 + p = \rho_1 h_1 \gamma_1^2 v_1^2 + p_1 \ \ \ \text{(momentum)} \label{momentum} \\
			\rho (h \gamma - 1) \gamma vs. = \rho_1 (h_1 \gamma_1 - 1) \gamma_1 v_1  \ \ \ \text{(energy)} \label{energy}.
		\end{gather}
	\end{subequations}
	{where the} fluid downstream of the shock is represented by the subscript `1'; no subscript refers to the upstream fluid.
	
	The conservation of energy expression in Equation \eqref{energy} is related to the rate of energy input by the jet, $Q$, by multiplying by the cross-sectional area of the jet {({cf.} \citep{Walg+2013}, specifically their Equation (26))}. That is:
	\begin{equation}
		Q = \rho_j (h_j \gamma_j - 1) \gamma_j v_j c^2 \Omega r^2 ,
		\label{jetcons}
	\end{equation}
	where the factor of $c^2$ is added to convert the dimensionless enthalpy, $h_j$, to the specific enthalpy, $v_j$ is the bulk velocity of the jet plasma, and $\gamma_j$ is the corresponding Lorentz factor.
	We can, therefore, obtain an expression for the density of the jet plasma some distance $r$ along the jet in terms of the dimensions and energetics of the jet as follows:
	\begin{equation}
		{\rho}_j(r) = \frac{Q}{\gamma_j {v}_j c^2 (h_j {\gamma}_j - 1) \Omega r^2} ,
		\label{jetdensity}
	\end{equation}
	where the density at the jet head, $\rho_j \equiv \rho_j(R_s)$, is of particular interest for the radio source dynamics.
	
	We now derive the Rankine--Hugoniot jump conditions relating the density and velocity of both the jet plasma and the ambient medium. The bulk velocity of the ambient medium in the observer frame is zero at all times for random particle motions. As a result, the bulk velocity of these particles in the frame of the shock front, $v_1$, is exactly equal to the expansion rate of the shock in the observer frame, $v_s$, i.e., $v_1 \equiv -v_s$. 
	By contrast, the bulk velocity of the upstream fluid particles in the jet is non-zero, defined as ${v}_j$ in the observer frame. 
	Following \citep{Turner+2023}, the conservation of momentum flux equation can, therefore, be rewritten as:
	\begin{equation}
		{\rho}_j h_j {\gamma}_j^2 \gamma_s^2 ({v}_j - v_s)^2 = \rho_x h_x \gamma_s^2 v_s^2 ,
		\label{jethead}
	\end{equation}
	where $h_j$ is the dimensionless specific enthalpy of the jet, and $\rho_x$ and $h_x$ are the density and dimensionless specific enthalpy of the (external) ambient medium, respectively. 
	Rearranging yields a relationship between the jet-head advance speed and the bulk velocity of the jet {(\mbox{{cf.} \citep{Marti+1994, Marti+1997, Rosen+1999}}, and subsequent authors)}:

	\begin{equation}
		v_s \equiv \frac{dR_s}{dt} = \frac{{v}_j}{1 + [{\rho}_j h_j {\gamma}_j^2/(\rho_x h_x)]^{-1/2}} ,
		\label{vs}
	\end{equation}
	where the dimensionless quantity $\eta_R = {\rho}_j h_j {\gamma}_j^2/(\rho_x h_x)$ is a function of properties of the jet and ambient medium. That is:
	\begin{equation}
		\eta_R(r) = \frac{Q h_j {\gamma}_j}{k h_x {v}_j c^2 (h_j {\gamma}_j - 1) \Omega r^{2 - \beta}} .
		\label{L}
	\end{equation}
	where we have made use of the power-law approximation for the local density of the ambient medium, $\rho = kr^{-\beta}$, and Equation \eqref{jetdensity} for the jet plasma density.
	

	The jet length is found by integrating Equation \eqref{vs} with respect to time; however, an analytical solution is only possible in the limits $\eta_R \rightarrow 0$ and $\eta_R \rightarrow \infty$  ({e.g.,} 
	\citep{Matzner+2003, Bromberg+2011}). \citet{Turner+2023} solved this integral numerically using a fourth-order Runge--Kutta method on the following system of three ordinary differential equations:
	\begin{equation}
		\begin{split}
			\dot{R}_s &= v_s \\
			\dot{v}_s &= \frac{(\beta - 2) {v}_j v_s}{2 R_s \eta_R^{1/2} [1 + \eta_R^{-1/2}]^2} \\
			\dot{\gamma}_s &= \frac{\gamma_s^3 v_s \dot{v}_s}{c^2} .
		\end{split}
		\label{jet system}
	\end{equation}
	{We} note that in the interests of clarity, we have omitted the transverse density and velocity structures of the flow along the jet from the above derivation; we refer the reader to \mbox{Section 2.2.2} of \citet{Turner+2023} for a complete description.

	\subsubsection{Lobe Formation}
	\label{sec:Lobe formation}
	
	The energy supplied by the central nucleus is initially focussed over a small range of angles within the half-opening angle of the jet. Beyond some lobe formation length scale, the energy must be distributed across the $2\pi$ steradians of the shocked shell. The source expansion in these two phases is described by the differential equations for the relativistic jet (Section \ref{sec:Relativistic hydrodynamic equations}) and lobe and shocked shell (Section \ref{sec:RAiSE angles}). 
	\citet{Turner+2023} combined these frameworks by modelling the expansion of the radio source as a two-phase fluid, where each angular volume element is assumed to comprise a fraction $\Lambda(t)$ of lobe plasma at any given time $t$. 
	\citet{Turner+2023} related the acceleration of the ellipsoidal bow shock surrounding the lobe to the acceleration in the jet- and lobe-dominated expansion phases, $\dot{v}_{s,j\!\;\!e\!\;\!t}$ (Equation \eqref{jet system}) and $\dot{v}_{s,l\!\;\!o\!\;\!b\!\;\!e}$ (Equation \eqref{supersonic system}), respectively, as follows:
	\begin{equation}
		\dot{v}_s(\theta) = [1 - \Lambda] \dot{v}_{s,j\!\;\!e\!\;\!t} \eta(\theta) + \Lambda \dot{v}_{s,l\!\;\!o\!\;\!b\!\;\!e} (\theta) ,
		\label{twophase}
	\end{equation}
	where $\Lambda$ is the fractional contribution of the lobe plasma to the acceleration of the bow shock at a given time. The other two coupled ordinary differential Equations (for the velocity and derivative of the Lorentz factor) are identical for both fluids, and thus, do not require any modification.
	
	\citet{Turner+2023} defined the transition from a jet-dominated to a lobe-dominated flow based on the length scale at which lobe formation commences. This length scale is calculated by equating the densities of the jet plasma and ambient medium (e.g., \citep{Alexander+2006, Krause+2012}). \citet{Turner+2023} parametrised the transition from jet- to lobe-dominated expansion by using the ratio of these densities:
	\begin{equation}
		\mathcal{L}(t) = \frac{{\rho}_j}{\rho_x} = \ \ \frac{\eta_R(R_s(\theta=0, t))}{{\gamma}_j^2}
		\label{bigL}
	\end{equation}
	where $\eta_R(r)$ is defined in Equation \eqref{L} and is evaluated for the length of the jet at time $t$. 
	\citet{Turner+2023} used this ratio to calculate the fractional contribution of the lobe to source expansion:
	\begin{equation}
		\Lambda(t) = e^{-\mathcal{L}^2(t)/(2 \log 2)} ,
		\label{lambda}
	\end{equation}
	where $\Lambda(t) \rightarrow 0$ in the jet-dominated expansion phase and $\Lambda(t) \rightarrow 1$ in the lobe-dominated phase.
	
	This two-phase fluid model describes the evolution of the bow shock across the transition from a jet-dominated to lobe-dominated flow. A more complete description requires consideration of lobe formation inside the shock front. We refer the interested reader to Section 2.4.2 of \citet{Turner+2023} for these details.
	
	\subsubsection{Thermal Pressure}
	\label{sec:Thermal pressure}
	
	The \citet{Turner+2023} relativistic jet model (Section \ref{sec:Relativistic hydrodynamic equations}) and their earlier lobe-dominated expansion model (Section \ref{sec:RAiSE angles}) derive the jet and lobe length evolution by considering conservation of momentum flux (Equation \ref{jethead}); however, the relative magnitudes of the ram and thermal pressure components after the interaction are not explicitly calculated. 
	These pressure components are difficult to separate directly using the conservation equations; however, we know the lobe evolution is driven entirely by the thermal component in the limit $t\rightarrow \infty$. 
	\citet{Turner+2023}, therefore, found the thermal pressure at earlier times by iteratively solving (backwards in time) the relevant differential equations with the initial condition at $t\rightarrow \infty$. We refer the interested reader to Section 2.3.4 of \citet{Turner+2023} for a complete description of the separation of the ram and thermal components of the lobe internal pressure.

	\section{Discussion}
	\label{sec:Discussion}
	
	In preceding sections, we have presented the theory underpinning the key classes of analytical models describing the dynamics of kiloparsec-scale radio AGN jets and lobes. The same physical principles are considered in each of these models, notably ram pressure against the ambient medium and an adiabatic equation of state; however, their implementation between model classes differs greatly---as we discuss in Section \ref{sec:Common physical assumptions}. We compare the accuracy of predictions for each model type relative to the outputs of a three-dimensional relativistic hydrodynamic simulation in Section \ref{sec:Comparison to hydrodynamic simulations}. We then assess for which regions of parameter space the different model classes yield comparable results, and conversely, those where large differences are expected, in Section \ref{sec:Parameter space exploration}.
	
	\subsection{Similarity of Key Model Classes}
	\label{sec:Common physical assumptions}
	
	The four key classes of analytical models examined in this review share common physical principles to explain the dynamics of kiloparsec-scale radio sources (see Table \ref{tab:tab1}). \citet{Scheuer+1974} modelled the forward expansion of the source based on the momentum flux of the jet and invokes internal energy to calculate the sidewards expansion of the lobe. \mbox{\citet{Falle+1991}}, instead, modelled the forwards expansion by considering the adiabatic expansion of the lobe due to an increase in internal energy while using the jet momentum flux to relate the shape of the lobe to the opening angle of the jet. Meanwhile, the \mbox{\citet{Hardcastle+2018}} and \citet{Turner+2023} models smoothly transition their dynamics between the jet- and lobe-dominated expansion phases predicted by these earlier models. 
	
	\begin{table}[H]  \tablesize{\scriptsize}
		\caption{Summary of the model assumptions for each of the four key classes of analytical models considered in this review. First column: analytical model. Second column: type of solution (analytical or semi-analytic), together with a note on the assumed ambient gas density profile. The third and fourth column: physical process driving the jet and lobe length expansion at early and late times, respectively. The fundamental equations used to derive the source evolution, and the time dependence of the length expansion, are also listed.}
		\renewcommand{\arraystretch}{1}
		\newcolumntype{C}{>{\centering\arraybackslash}X}
		\begin{tabularx}{\textwidth}{CCCC}
			\toprule
			\textbf{Model}	& \textbf{Type}	& \textbf{Early-Time Evolution} & \textbf{Late-Time Evolution} 	\\
			\midrule
			{Scheuer} 
			(1974;\par
			Model~A)  & analytical\par \textit{(constant density)}	& momentum flux\par \textit{(Bernoulli equation)}\par\footnotesize $R(t) \propto t^{2/(4 - \beta)}$ & --		\\
			{Falle} (1991) 	& analytical\par \textit{(power-law density profile)} & -- & internal pressure\par \textit{(first law of thermodynamics)}\par\footnotesize $R(t) \propto t^{3/(5 - \beta)}$	\\
			{Hardcastle} (2018)  	& semi-analytic\par \textit{(spherically symmetric density profile)}	& momentum flux \textit{(non-relativistic shock-jump conditions)}\par\footnotesize $R(t) \propto t^{2/(4 - \beta)}$ & internal pressure\par \textit{(non-relativistic shock-jump conditions)}\par\footnotesize $R(t) \propto t^{3/(5 - \beta)}$ 	\\
			{Turner} et al. (2023;\par
			RAiSE)  & semi-analytic\par \textit{(spherically symmetric density profile)}	& momentum flux\par \textit{(relativistic hydrodynamics)}\par\footnotesize  $R(t) \propto t$ $^\dagger$ & internal pressure\par \textit{(first law of thermodynamics)}\par\footnotesize $R(t) \propto t^{3/(5 - \beta)}$ 	\\
			\bottomrule
		\end{tabularx}
		\noindent{\footnotesize{$\dagger$ Solution for jet-head length expansion prior to lobe formation length scale.}}
		\label{tab:tab1}
	\end{table}
	\vspace{-15pt}
	\subsubsection{Early-Time Evolution}
	
	The source length evolution in the \citet{Scheuer+1974} model, and the (jet-dominated) early-time expansion phases of the \citet{Hardcastle+2018} and \citet{Turner+2023} models, are derived considering the relative amplitudes of the momentum flux of the jet and the thermal pressure of the ambient medium (Equation \eqref{momentum}). The relativistic hydrodynamic equations used in the theory of \citet{Turner+2023} can be simplified to obtain the expressions found by the \citet{Scheuer+1974} class of models. In particular, the jet-head advance speed derived in Equation \eqref{vs} is integrated to yield the source length as follows:
	\begin{equation}
		\begin{split}
			R(t) &= \int_0^t \frac{dR_s}{dt} dt \\
			&\approx \left(\frac{Q h_j {\gamma}_j v_j}{\Omega k h_x c^2 (h_j {\gamma}_j - 1)}\right)^{1/(4-\beta)} \left(\frac{(4-\beta)t}{2} \right)^{2/(4-\beta)} ,
		\end{split}
	\end{equation}
	where the second equality is valid for $\eta_R(r) \ll 1$, which coincides with the formation of lobes on a length scale of order 1\,kpc \citep{Turner+2023}. This equation converges to that proposed by \citet{Rees+1971} and \citet{Scheuer+1974} (their Model A) by taking their limit of massless particles with velocity $c$ (i.e., $h_{j,x} \rightarrow 1$ and $\gamma_j \rightarrow \infty$):
	\begin{equation}
		R_s(t) = \left(\frac{Q}{\Omega k c}\right)^{1/(4-\beta)} \left(\frac{(4-\beta)t}{2} \right)^{2/(4-\beta)} ,
	\end{equation}
	where for their assumption of a uniform ambient medium, we set $\beta = 0$ and $k = \rho$.
	
	The similarity of the early-time evolution predicted by the \citet{Hardcastle+2018} model and \citet{Scheuer+1974} class of models is immediately apparent by comparing their expression for the jet-head pressure (Equation \eqref{hardcastle pressure}). In the limit $t \rightarrow 0$, their expression converges to the ram pressure component as {follows}: 
\begin{equation}
\begin{split}
    p_s(\theta=0) &= \frac{\varepsilon Q R_s(\theta=0)}{2cV} \ \ \ \text{(Hardcastle)} \\
    &= \frac{\kappa_1 Q }{\Omega R_s^2(\theta = 0) c} \ \ \ \text{(Scheuer)},
\end{split}
\end{equation}
	where the geometric factor assumed by \citet{Hardcastle+2018} is identified as $\varepsilon = 2\kappa_1/\Omega$, using the terminology of \citet{Scheuer+1974}; in other words, this geometric factor largely corresponds to the solid angle of the jet and the jet-head region. \citet{Hardcastle+2018} derived the jet-head advance speed using the non-relativistic Rankine--Hugoniot shock jump conditions (Equation \eqref{hardcastle system}), largely equivalent to the ram pressure argument employed by \citet{Scheuer+1974}. Because of this, the source length evolution predicted by the \citet{Hardcastle+2018} model early in the source lifetime matches that of the \citet{Scheuer+1974} class of models for the same input parameters, i.e., $R \propto t^{1/2}$ for a constant density ambient medium. By contrast, the jet-dominated expansion phase of the \citet{Turner+2023} model yields a very different evolutionary history to these two model types, with $R \propto t$ prior to lobe formation, as those authors do not take the non-relativistic limit of the hydrodynamic equations.
	
	\subsubsection{Late-Time Evolution}
	
	The lobe length expansion in the \citet{Falle+1991} class of models and the (lobe-dominated) late-time expansion phases of the \citet{Hardcastle+2018} and \citet{Turner+2023} models are calculated by considering the internal energy of the relativistic lobe plasma evolving under an adiabatic equation of state. The first law of thermodynamics (Equation \eqref{gov}) applied to a self-similar ellipsoidal shell yields lobe length growth of the form $R \propto t^{3/5}$ (\mbox{Equation \eqref{shockradius}}, constant gas density form is stated here for simplicity), consistent with findings for supernova remnants. The \citet{Falle+1991} class of models use this dependence directly, while the lobe-dominated expansion phase of the \citet{Turner+2023} model also permits an evolving lobe axis ratio due to a non-power law ambient medium. The late-time evolution of the \citet{Turner+2015} model (and subsequent versions of RAiSE) ultimately transitions from the supersonic to subsonic regime as the jet-head advance speed slows ($R \propto t^{1/3}$). This is not critical when modelling powerful lobed radio sources which drive strong shocks---the focus of this review---but is essential when considering the coasting (inactive) phase of remnant radio sources.
	
	\citet{Hardcastle+2018} did not make any explicit assumptions about the lobe axis ratio; however, in the limit $t \rightarrow \infty$, their expressions for the expansion rate along both the major and minor axes converge, leading to a spherical lobe (at least in the special case of a constant density ambient medium). The expression for their growth rate at late times (Equation \eqref{hardcastle system}) becomes:
	\begin{equation}
		\frac{dR_s}{dt} = \sqrt{\frac{(\Gamma_s + 1)([\xi\Gamma_c + (1 - \xi)\Gamma_s - 1]Qt + f(N,T,v_s))}{2\rho_x V_s}} ,
	\end{equation}
	where we have assumed that the pressure along both axes is dominated by the internal energy component (Equation \eqref{hardcastle pressure}) and that the sound speed in the shocked gas is comparable to that of the ambient medium (i.e., $c_s \sim \sqrt{\Gamma_s p_x/\rho_x}$), and applied the relationship for the ratio of the lobe and shocked shell volumes (Equation \eqref{Vh}). This differential equation again yields a solution of the form $R \propto t^{3/5}$ for a constant density ambient medium, if the internal energy contribution from the shocked gas, $f(N,T,v_s)$, is ignored. Such an assumption is reasonable for high jet powers or moderate ambient densities; however, the late-time evolution diverges significantly from this prediction for weak jets or high density environments---we examine this point further in Section \ref{sec:Accuracy of analytical models}. 
	
	\subsubsection{Source Morphology}
	
	The lobe volume and, hence, axis ratio are calculated using the remaining equations not previously invoked in the calculation of the source length expansion. For example, in the \citet{Scheuer+1974} model, source expansion along the jet axis is modelled based on ram pressure arguments, while the sidewards expansion is derived considering the internal energy of the injected lobe plasma; \citet{Hardcastle+2018} made similar arguments, while \citet{Turner+2023} invoked internal energy to describe the formation and inflation of their lobe within the confines of a surrounding shocked gas shell. By contrast, the self-similar expansion assumed in the \citet{Falle+1991} class of models, and the lobe-dominated expansion phase of the \citet{Turner+2023} model, implicitly sets the lobe volume based on its length, which is calculated from internal energy. These models use ram pressure arguments to relate the internal conditions of the lobe to those of the ambient medium (Equation \eqref{pressratio}), and to relate the jet half-opening angle to the source axis ratio.

	\subsection{Comparison to Hydrodynamic Simulations}
	\label{sec:Comparison to hydrodynamic simulations}
	
	To test the analytical models, we compare the dynamics of the four key model classes introduced above to the results of hydrodynamic simulations run using the PLUTO code~\citep{Mignone+2007}. Below, we describe the existing simulations run by \citet{Yates+2022} for powerful FR-II radio galaxies
	and test the analytical models by comparing the predicted evolution of the source length, lobe axis ratio, and jet-head pressure throughout the source lifetime to the simulation results. 
	
	\subsubsection{Hydrodynamic Simulation Dynamics}
	\label{sec:Hydrodynamic simulation dynamics}
	
	The hydrodynamic simulations of \citet{Yates+2022} consider an initially conical jet of half-opening angle $\theta_{j} = 10$ degrees and bulk flow with Lorentz factor $\gamma_j = 5$. Their high-powered jet ($Q = 3 \times 10^{38}$\,W) expands into a spherically symmetric King profile with a core density of $\rho_c = 2.41\times10^{-24}$\,kg\,m$^{-3}$, core radius of $r_c = 144$\,kpc, and slope described by the coefficient $\beta' = 0.38$ (for details, see \citep{Yates+2021}). Those authors considered both jets located at the centre of the gas distribution, as well as offset jets. In this review, we will only consider their cluster-centred jet simulation, as the theory underpinning the analytical models assumes a spherically symmetric environment. Their simulations resulted in lobed \citet{Fanaroff+1974} Type-II sources forming on a timescale of $\sim$1\,Myr, and considered the late-time evolution up to an age of 35.1\,Myr. 
	
	We extract time series for the evolution of the source length, the lobe axis ratio, and the jet-head pressure from the hydrodynamic simulation outputs (for details, see \citep{Turner+2023}). These are critical dynamical quantities in calculation of both the source evolution and the radio-frequency synchrotron emission, and thus, should be considered to assess the accuracy of the analytical models. The potentially more informative lobe volume and volume-weighted pressure are poorly constrained prior to lobe formation as large regions near the core remain partially occupied by ambient gas; the calculation of these quantities in the hydrodynamical simulation is highly dependent on the threshold used to separate ambient gas from the jet plasma.

	\subsubsection{Accuracy of Analytical Models}
	\label{sec:Accuracy of analytical models}
	
	The critical intrinsic parameters characterising radio source evolution in analytical models are the jet kinetic power, source age, and properties of the ambient medium. The spherically symmetric King profile used by \citet{Yates+2021} for the ambient medium in their simulations is readily modelled by both the \citet{Hardcastle+2018} and \citet{Turner+2023} models, but not the older models. The original form of the \citet{Scheuer+1974} model assumes a constant density ambient medium, while the \citet{Falle+1991} model employs a slightly more general power-law of the form $\rho = k r^{-\beta}$; in Section \ref{sec:The First Models}, we similarly derived the \mbox{\citet{Scheuer+1974}} model for a power-law gas density profile. To facilitate a meaningful comparison between all four model classes, we modify the \citet{Falle+1991} and \citet{Scheuer+1974} model following the approach of \citet{Turner+2015}, by approximating the gas density profile as a series of contiguous power-laws with piecewise solutions. The implementation of these models for a general ambient medium is available on our \texttt{GitHub} online {repository}\endnote[1]{\customlabel{note1}{1}\url{https://github.com/rossjturner/analytical_models}}.

	\paragraph{{Source} Length} 
	
\begin{figure}[H]
		\begin{adjustwidth}{-\extralength}{0cm}\centering
			
			\includegraphics[width=17cm,trim={100 87.5 112.5 115},clip]{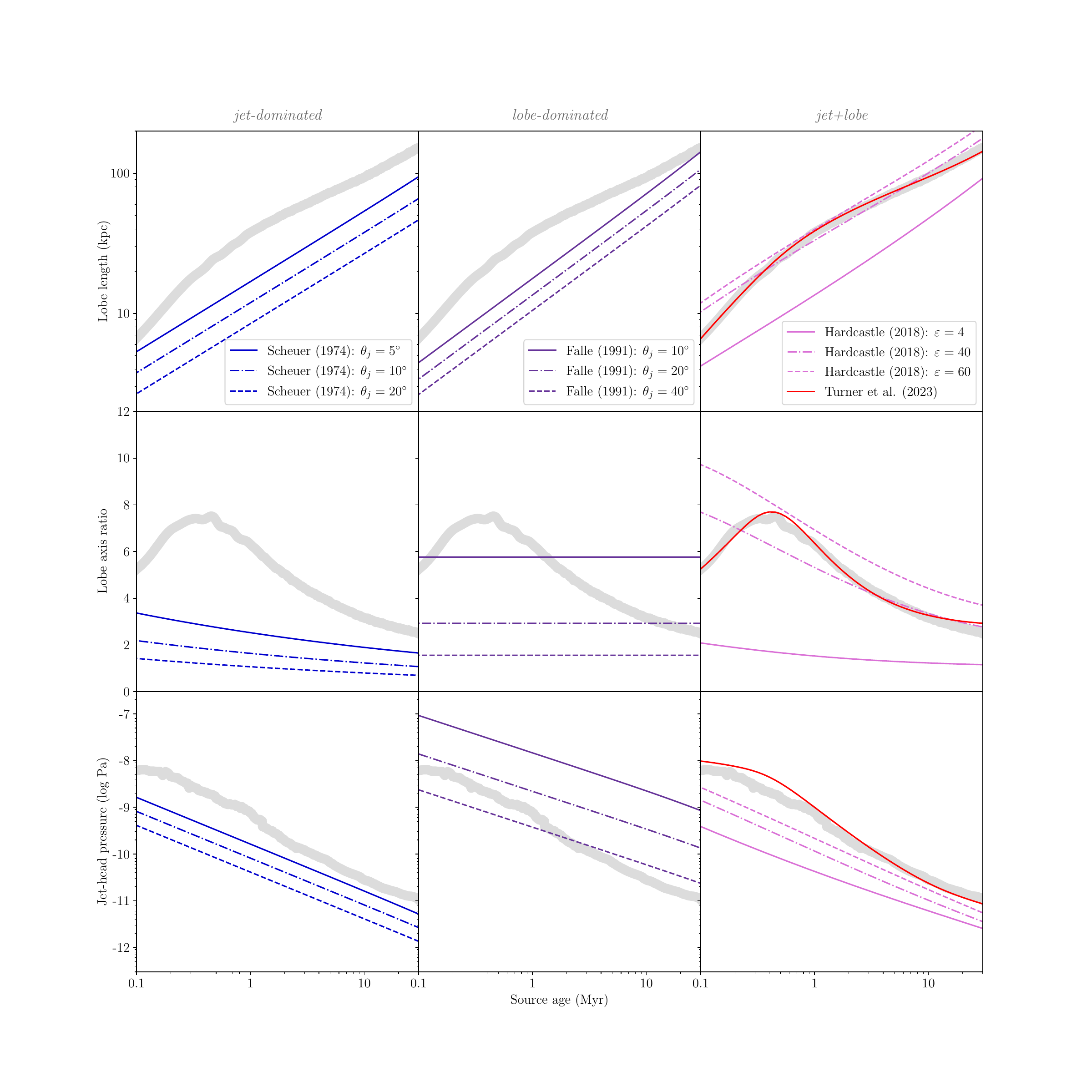}
		\end{adjustwidth}
		\caption{{Comparison} 
			of the four classes of analytical models to the hydrodynamic simulation (grey shading) used by \citet{Turner+2023} (specifically, their Figure 3) to assess the success of their model near the commencement of lobe formation. Top panel: source length evolution. Middle panel: lobe axis ratio. Bottom panel: jet head pressure. The \citet{Scheuer+1974} and \citet{Falle+1991} class models are shown for a range of jet half-opening angles $\theta_j$, while the \citet{Hardcastle+2018} model is shown for a range of jet-head cross-sections $\varepsilon$. The RAiSE model \citep{Turner+2023} is shown for its optimised set of parameters.}
		\label{fig:turner_hyd}
\vspace{-5pt}
	\end{figure}

\begin{figure}[H]
		\begin{adjustwidth}{-\extralength}{0cm}\centering
			\includegraphics[width=\fulllength,trim={97.5 60 112.5 87.5},clip]{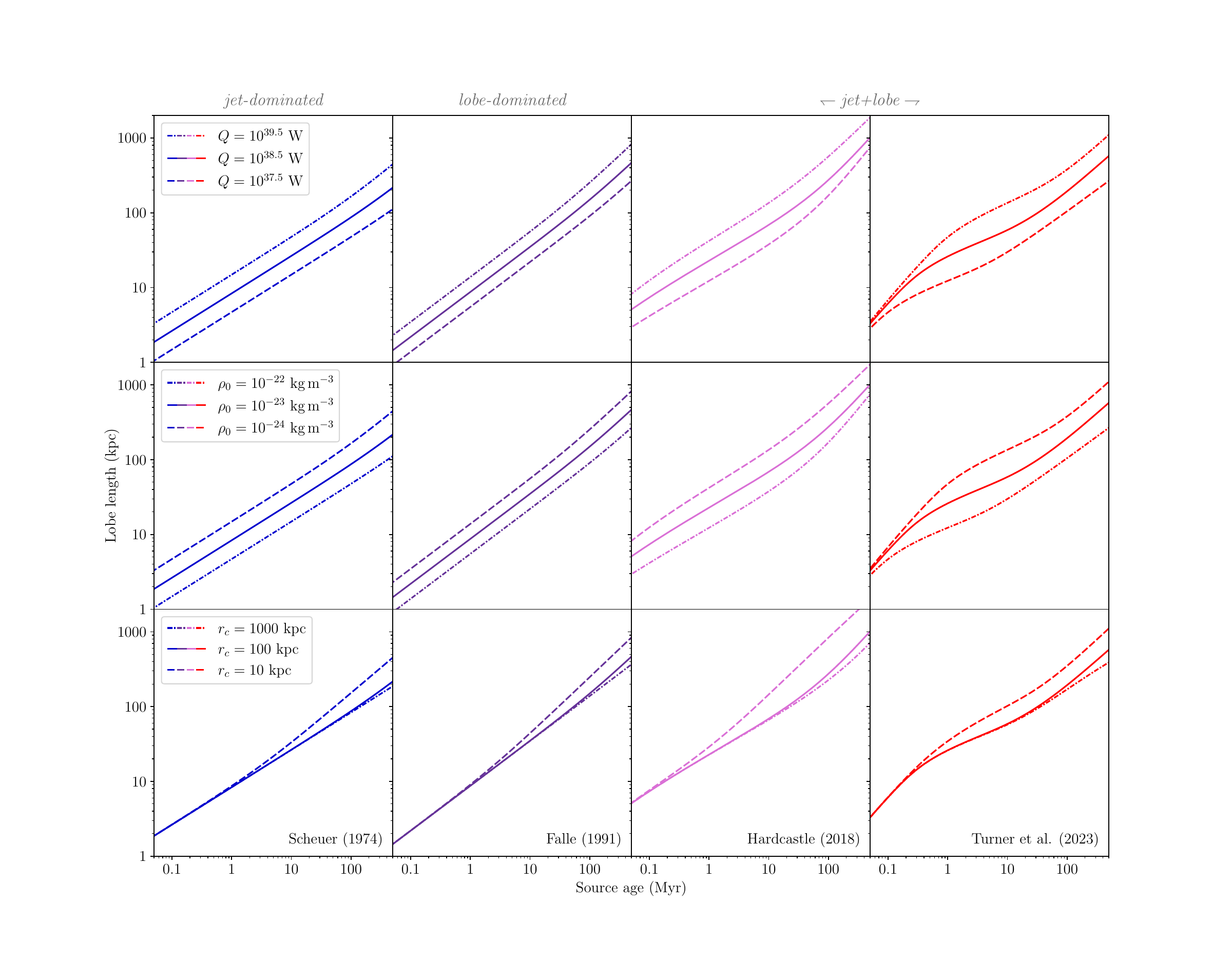}
		\end{adjustwidth}
		\caption{{Source} 
			length evolution predicted by the four classes of analytical models for a range of input intrinsic parameters. The evolutionary tracks for the base set of parameters, i.e., $Q = 3 \times 10^{38}$\,W, $\rho_0 = 10^{-23}$\,kg\,m$^{-3}$ and $r_c = 100$\,kpc, are shown as solid lines. The source evolution for either a lower ($Q = 3 \times 10^{37}$\,W) or higher ($Q = 3 \times 10^{39}$\,W) jet power is shown in the top panel with dashed and dot-dashed lines, respectively. The middle panel shows evolutionary tracks with lower ($\rho_0 = 10^{-24}$\,kg\,m$^{-3}$) or higher ($\rho_0 = 10^{-24}$\,kg\,m$^{-3}$) core densities. The bottom panel plots the evolution for either a lower ($r_c = 10$\,kpc) or higher ($r_c = 1000$\,kpc) scale radius of the ambient gas density profile.}
		\label{fig:length_comp}
	\end{figure}

	The source length evolution for the four analytical models is shown in the top panel of Figure \ref{fig:turner_hyd}. The jet power, source age, and ambient gas density profile are in all cases identical to the inputs to the hydrodynamic simulation of \citet{Yates+2021}; however, some of the more minor model parameters are varied to obtain the best representation of each model class. Specifically, the \citet{Scheuer+1974} and \citet{Falle+1991} model evolutionary tracks are shown for three plausible values of the jet half-opening angle $\theta_j$, while the \mbox{\citet{Hardcastle+2018}} model is shown for three values of their equivalent geometric factor $\varepsilon$. The free parameters in the \citet{Turner+2023} model have previously been calibrated based on this hydrodynamic simulation, and thus, results for only a single set of parameters for this model are shown. 
	The resulting evolutionary tracks are consistent with the discussion in \mbox{Section \ref{sec:Common physical assumptions}}: the \citet{Scheuer+1974} and (jet-dominated) early-time \citet{Hardcastle+2018} models follow an approximately $R \propto t^{1/2}$ growth rate (the dependence expected for a flat atmosphere), while the \citet{Falle+1991} and (lobe-dominated) late-time \citet{Hardcastle+2018} models follow an $R \propto t^{3/5}$ expansion rate (also see Figure \ref{fig:length_comp}). By contrast, the relativistic hydrodynamic equations used by \citet{Turner+2023} yield quite different evolutionary tracks in the jet-dominated expansion phase, and are more consistent with the hydrodynamic simulation. At late times ($>10$\,Myr), the \citet{Turner+2023} model predicts slower expansion close to an $R \propto t^{1/2}$ relationship, converging towards the same limit as the other models at later times. We explore this in more detail in Section \ref{sec:Parameter space exploration}.

	\paragraph{{Lobe} Axis Ratio}
	
	The evolution of the lobe axis ratio is shown in the middle panel of Figure \ref{fig:turner_hyd}. The \citet{Turner+2023} model captures axis ratio evolution prior to lobe formation, as it considers a two-phase fluid with an initially low, but non-zero, fraction of jet plasma in the region between the jet and bow shock. By contrast, the other analytical models initially disagree with the hydrodynamic simulation, as they assume a plasma-filled lobe structure throughout the source lifetime. Upon lobe formation, the \citet{Hardcastle+2018} model agrees well with the hydrodynamic simulation assuming a compact jet-head region with a geometric factor $\varepsilon \gtrsim 40$ (corresponding to a jet head of radius 12.6\,kpc for a typical 100\,kpc jet). \mbox{The \citet{Falle+1991}} class of models, of course, yields a constant lobe axis ratio (which is why these models are often referred to as ``self-similar'') throughout the source evolutionary history, while the \citet{Scheuer+1974} model predicts a rapidly decreasing axis ratio with the lobe approaching a near-spherical shape in the steeper sections of the ambient gas density profile ($A \propto t^{(\beta -2)/(16 - 4\beta)}$; see Section \ref{sec:The First Models}). These last two models do not explicitly separate the lobe and shell material, so for a fairer comparison, we are guided by the numerical results of \citet{Turner+2020a} in assuming that the lobe axis ratio scales to that of the shell as $A = A_s^{1.7}$. 
	
	\paragraph{{Jet-Head} Pressure}
	
	We finally compare the predicted evolution of the jet-head pressure to that measured from the hydrodynamic simulation (Figure \ref{fig:turner_hyd}, bottom panel). The majority of the evolutionary history probed by the hydrodynamic simulation ($\lesssim$10\,Myr) is associated with a significant ram pressure component at the jet head, in addition to a thermal component that scales approximately in proportion to the ram pressure component. The \citet{Scheuer+1974} and \citet{Hardcastle+2018} models directly consider the ram pressure component, and hence, predict a power-law evolution of jet head pressure ($p \propto t^{-1}$ for a constant gas density ambient medium; see Sections \ref{sec:The First Models} and \ref{sec:Hertfordshire model}), which is broadly similar to the thermal jet head pressure component measured by the hydrodynamic simulation. The self-similar expansion model of \citet{Falle+1991} yields a flatter relationship of the form $p \propto t^{-4/5}$. \citet{Turner+2023}, instead, explicitly modeled both the thermal and ram pressure components throughout the source evolutionary history; this model accurately captures the steepening in the rate of change of jet-head pressure during the transition between these two limiting cases.

	\subsection{Parameter Space Exploration}
	\label{sec:Parameter space exploration}
	
	In this section, we compare the consistency of observable predictions between the four model classes. We select three parameters for our comparison. Source length and axis ratio are directly measurable model predictions, while synchrotron luminosity integrated over the entire lobe is a quantity which can be approximated from source dynamics---noting that detailed consideration of particle acceleration and loss processes is required for a full calculation (e.g., \citep{KDA+1997}). These radio source attributes are critical for estimating the jet energy budget through a parameter inversion of observables \citep{Shabala+2008, Turner+2015}. 
	
	We investigate the behaviour of the different model classes for a range of input parameters, specifically the single-jet kinetic power $Q$, core density $\rho_0$, and scale radius $r_c$ of the ambient gas density profile. The base-case set of parameters is chosen to match the hydrodynamic simulation of \citet{Yates+2021}, but with $\rho_0 = 10^{-23}$\,kg\,m$^{-3}$ and $r_c = 100$\,kpc. The jet half-opening angles for the \citet{Scheuer+1974} and \citet{Falle+1991} models are set to $\theta_j = 10^\circ$ and $\theta_j = 20^\circ$, respectively, as these closely match the hydrodynamic simulation evolution for the three key dynamical parameters of source length, lobe axis ratio, and jet-head pressure (cf. Figure \ref{fig:turner_hyd}). For the same reasons, we set the geometric factor to $\varepsilon = 40$ in the \citet{Hardcastle+2018} model.
	The evolutionary tracks for each model class are considered both for the base set of parameters and for a factor of ten variation to one of either the jet power ($Q = 3 \times 10^{37}$, $3 \times 10^{38}$ or $3 \times 10^{39}$\,W), core density ($\rho_0 = 10^{-24}$, $10^{-23}$ or $10^{-22}$\,kg\,m$^{-3}$), or scale radius ($r_c = 10$, 100 or 1000\,kpc).
	
	\paragraph{{Source} Length}

	The source length evolution for the four model classes are shown in Figure \ref{fig:length_comp}. Changes in jet power and core density largely result in a constant offset (in logarithmic space) to the source length, consistent with the expected scalings in flat atmospheres of $R \propto Q^{1/4}$ and $R \propto \rho_0^{-1/4}$ for the jet-dominated \citep{Scheuer+1974} model, and $R \propto Q^{1/5}$ and $R \propto \rho_0^{-1/5}$ for the lobe-dominated \citep{Falle+1991} model class. The more sophisticated models of \citet{Hardcastle+2018} and \citet{Turner+2023} capture the transition between these limiting cases. For the same input parameters, the \citet{Hardcastle+2018} and \citet{Turner+2023} models predict similar dynamical evolution for the majority of simulated sources. However, the jet-dominant phase lasts longer for high-powered jets and in denser environments in the \citet{Turner+2023} model, making this model more sensitive to changes in these parameters than the other models. Variations to the scale radius of the ambient gas density profile yield qualitatively similar behaviour between the model classes, with faster expansion occurring for steeper atmospheres.
	
	\paragraph{{Lobe} Axis Ratio}

\begin{figure}[H]
		
		\begin{adjustwidth}{-\extralength}{0cm}\centering

			\includegraphics[width=17cm,trim={97.5 60 112.5 87.5},clip]{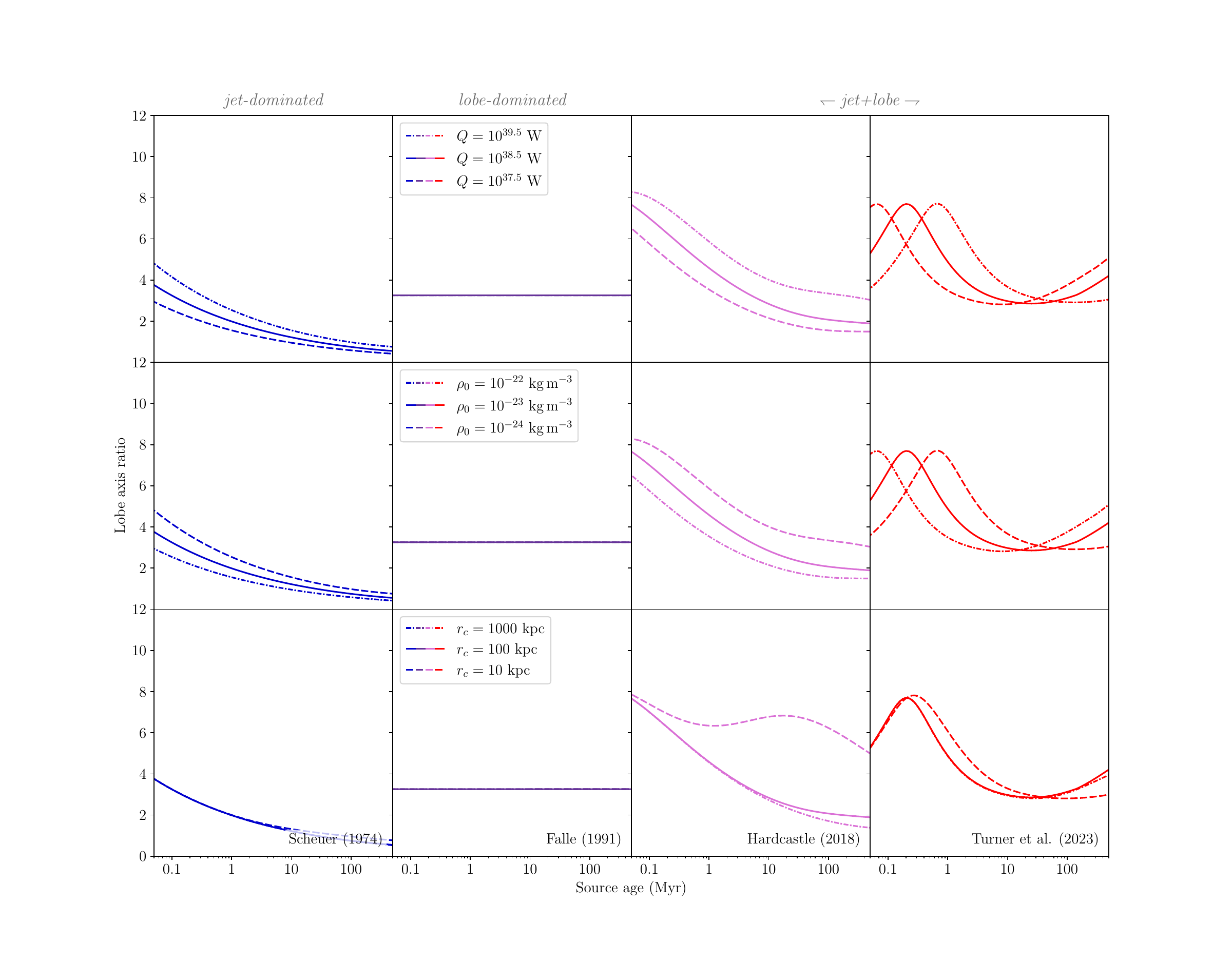}
		\end{adjustwidth}
		\caption[]{{Lobe}  
			axis ratio evolution predicted by the four classes of analytical models for a range of input intrinsic parameters. Panels and line styles are as in Figure \ref{fig:length_comp}.}
		\label{fig:axis_comp}
	\end{figure}

	The evolution in the lobe axis ratio is shown in Figure \ref{fig:axis_comp}. The \citet{Hardcastle+2018} and \citet{Turner+2023} models both predict the lobe axis ratio evolutionary tracks shift horizontally along the source age axis in response to variations in the jet power and core density, with lobe formation occurring earlier for less powerful jets and/or denser ambient media (e.g., see \citep{Turner+2023}). 
	The \citet{Scheuer+1974} model predicts a similar response to changes in the jet power and core density for the approximately constant ambient gas density section of the lobe axis ratio evolutionary tracks (i.e., $\lesssim$1\,Myr), but is inconsistent at later times once the ambient density profile begins to steepen (see Section \ref{sec:Comparison to hydrodynamic simulations}). By contrast, variations to the scale radius produce lobe axis ratio evolution that is largely inconsistent between the different model classes. The self-similar model of \citet{Falle+1991} yields a constant lobe axis ratio for all input parameters by construction. 
	
	\paragraph{{Synchrotron} Luminosity}

	Radio galaxies are detectable through their synchrotron emission. In analytical radio source models, this emission is typically calculated by assuming a scaling between the lobe pressure and magnetic field, acceleration of emitting particles at the jet termination shock, and subsequent losses due to adiabatic, synchrotron, and inverse-Compton losses due to upscattering of cosmic microwave background photons. 
	
	A full calculation of the radio-frequency luminosity of each lobe is beyond the scope of this review (see, e.g., \citep{KDA+1997,Shabala+2008,Turner+2015,Hardcastle+2018}). However, a useful estimate can be obtained by considering only adiabatic losses, which are directly related to the evolution of lobe pressure. The ``lossless'' luminosity calculated in this approach is related to the lobe volume and pressure as $L_\nu \propto p^{(\alpha + 3)/2}V$, where $\alpha \sim 0.7$ is the spectral index of the non-aged radio spectrum. The total internal energy of the lobe, $U \sim pV$, represents the fraction of the jet energy that is transferred to the lobe, and hence, is very similar in all models for a fixed time and jet power. In this case, the synchrotron luminosity scales with pressure as $L_\nu \propto p^{(\alpha + 1)/2} U$.
	
\begin{figure}[H]
		
		\begin{adjustwidth}{-\extralength}{0cm}\centering

			\includegraphics[width=17cm,trim={97.5 60 112.5 87.5},clip]{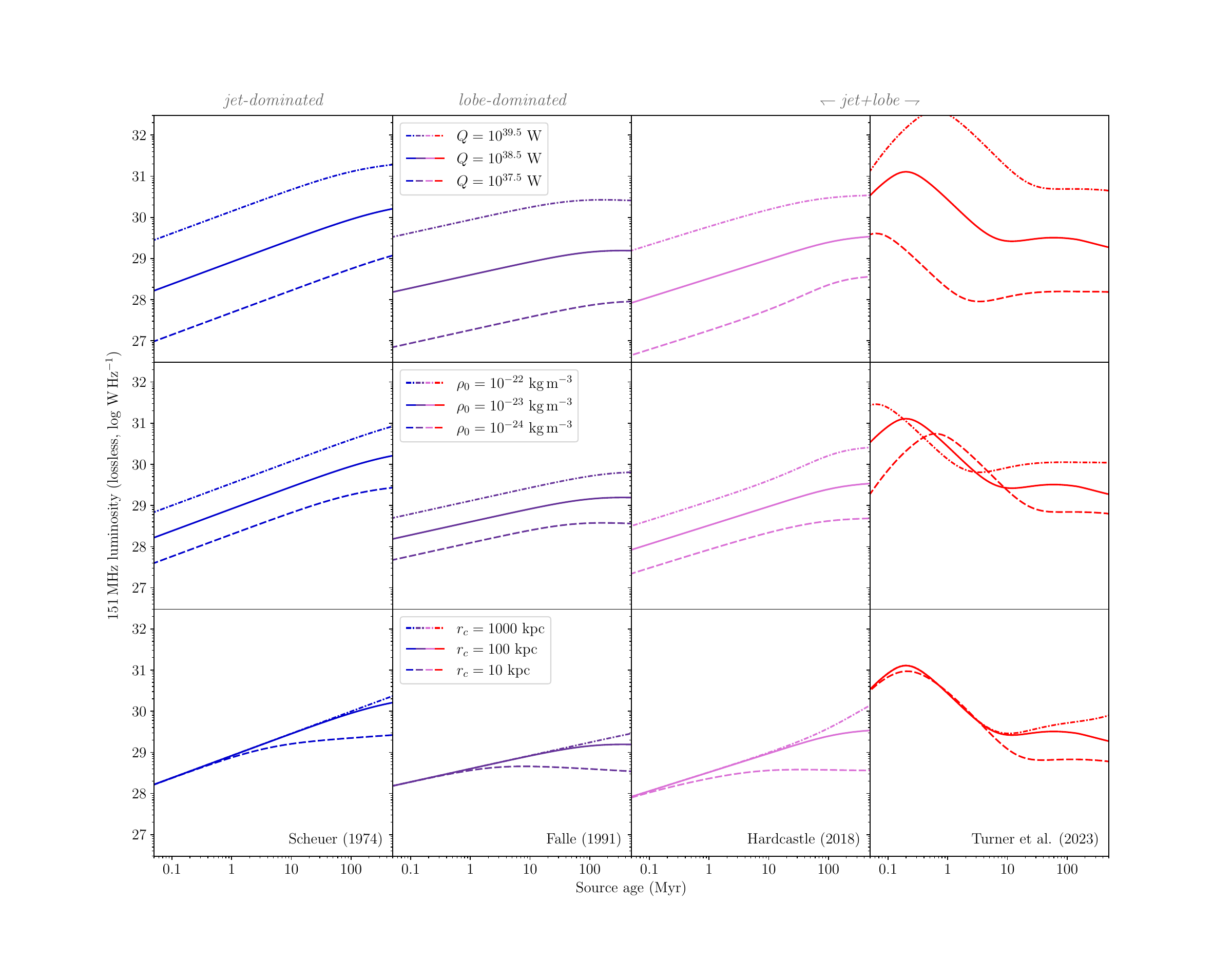}
		\end{adjustwidth}
		\caption{{Lossless}  
			synchrotron luminosity at 151\,MHz (see text for details). Panels and line styles are as in Figure \ref{fig:length_comp}.}
		\label{fig:pressure_comp}
	\end{figure}

	The evolution in the lossless synchrotron luminosity for each of the model classes is shown in Figure \ref{fig:pressure_comp} at a rest-frame frequency of 151\,MHz.
	The predictions of the \mbox{\citet{Scheuer+1974}} model are once again inconsistent with the later models, with radio luminosities up to a factor of 10 greater than the other models at late times.
	The remaining three models are consistent at late times---when their lobe pressures are largely derived based on changes in lobe internal energy---for all sets of input jet powers, core densities, and scale radii. The \citet{Falle+1991} and \citet{Hardcastle+2018} model classes also agree at early times, in contrast to the \citet{Turner+2023} model, which predicts higher luminosities as a result of significantly higher pressures prior to lobe formation when the system is dominated by the momentum flux of the relativistic jet. At these early times, the luminosities are a factor of 100-1000 higher than in the lobe-only models of \citet{Falle+1991} and \citet{Hardcastle+2018}.
	
	\section{Concluding Remarks}
	\label{sec:Conclusions}
	
	We have summarised and compared the main classes of analytical models describing the dynamics of kiloparsec-scale lobed radio galaxies. These models can be separated into two main classes, depending on whether the expansion of the radio source is driven by the momentum flux from the jet or by the internal lobe pressure. We presented the \citet{Scheuer+1974} and \citet{Falle+1991} models, respectively, to describe the general characteristics of other literature models in either of these two classes. We also examined separately the more recent models proposed by \citet{Hardcastle+2018} and \citet{Turner+2023}, which combine aspects of both jet momentum flux and lobe pressure.
	
	We compared the different model classes against each other, and with high-resolution hydrodynamic simulations, for a range of realistic input parameters. Our key findings are as follows:
	
	\begin{itemize}
		\item Jet momentum flux and lobe internal pressure dominate the early- and late-time radio source evolution, respectively. Both must be considered for a complete radio source model describing source dynamics after the lobe formation phase ($\sim$1\,Myr; \mbox{Section \ref{sec:Accuracy of analytical models}}).
		\item Realistic ambient gas density profiles (i.e., not constant or power-law) produce radio sources which are inconsistent with the self-similar lobe evolution predicted by the \mbox{\citet{Falle+1991}} class of models (Sections \ref{sec:RAiSE angles} and \ref{sec:Hertfordshire model}). This naturally explains the large axis ratios seen in giant radio galaxies.
		\item Relativistic jet dynamics is important for an accurate description of early source evolution, before the lobe formation phase (Section \ref{sec:Accuracy of analytical models}).
	\end{itemize}
	
	We make three of the four models considered in this review openly available. The code extending the \citet{Scheuer+1974} and \citet{Falle+1991} models to general atmospheres, and RAiSE (version 2023), are available in our \texttt{GitHub} online {repository}\endnotemark[\ref{note1}]$^{\!\;\!,}$\endnote[2]{\url{https://github.com/rossjturner/RAiSEHD}}. \citet{Hardcastle+2018} has also made their code available; we refer the interested reader to their paper.
	
	We conclude with a brief reflection on the next generation of analytical models.
	Existing analytical models neglect the interaction between the jet and the multi-phase interstellar medium of the host galaxy. Hydrodynamic simulations that model this interaction predict that the jet can spend $\gtrsim 1$\,Myr in the galaxy (e.g., \citep{Bicknell+2018}). The majority of observed radio sources are compact and short-lived \citep{Hardcastle+2019,Shabala+2020}, and hence, these processes are likely to be relevant to the bulk of the radio source populations.
	At the other end of the radio galaxy evolution scale, \citet{Hardcastle+2018} pointed out that for extreme losses, such as expected in large sources at high redshift, it is possible for the majority or even all of the jet energy to be radiated away. This mechanism can potentially limit the maximum size to which a radio source can grow. Existing analytical models decouple source dynamics from the synchrotron and inverse Compton radiative loss mechanisms, and hence, are not currently capable of tackling this issue. 
	
	\vspace{6pt} 
	
	\authorcontributions{
		Conceptualisation, R.J.T. and S.S.S.; methodology, R.J.T.; software, R.J.T.; validation, R.J.T.; 
		investigation, R.J.T.; resources, R.J.T.; writing---original draft preparation, R.J.T. and S.S.S.; writing---review and editing, R.J.T. and S.S.S.; visualisation, R.J.T. All authors have read and agreed to the published version of the manuscript.
	}
	
	\funding{
		This research received no external funding.}
	
	\dataavailability{{Not applicable.}}
	
	
	\conflictsofinterest{The authors declare no conflict of interest.} 
	
	
		
	
	\appendixtitles{yes} 
	\appendixstart
	\appendix
	\section[\appendixname~\thesection]{Early Jet--Lobe Models}
	\label{app:The First Models}
	
	%
	%
	We present a complete derivation for the lobe pressure and volume evolution of the \citet{Scheuer+1974} model assuming a power-law ambient density profile (rather than their assumed constant density medium), as outlined in Section \ref{sec:The First Models}.
	
	\subsection{Lobe Pressure}
	\label{app:Lobe pressure}
	
	The increase in the total energy of the cavity, $U$, over the time interval $\delta t$ due to the input kinetic energy $Q$ is given in Equation \eqref{energyeq}. \citet{Scheuer+1974} rewrote this first-order differential equation in terms of the jet length by defining a constant scaling between the lobe volume and this length of the form $V(R) = \kappa_2 R^\alpha$, where $\alpha, \kappa_2 > 0$ are constants. That is:
	\begin{equation}
		\delta U = \left[ Q \frac{dt}{dR} - \alpha \frac{U(\Gamma_c - 1)(q + 1)}{R} \right] \delta R ,
		\label{energyapp}
	\end{equation}
	where the time derivative of Equation \eqref{scheuer} gives the jet-head advance speed:
	\begin{equation}
		\begin{split}
			\frac{dR}{dt} &= \left(\frac{\kappa_1 Q}{\Omega k c}\right)^{1/(4-\beta)} \left(\frac{(4-\beta)t}{2} \right)^{(\beta -2)/(4-\beta)} \\
			& = \left(\frac{\kappa_1 Q}{\Omega k c}\right)^{1/2} R^{(\beta - 2)/2}.
		\end{split}
		\label{veloc}
	\end{equation}
	{The} solution to Equation \eqref{energyapp}, upon substituting the second expression above for the jet-head advance speed and assuming the initial condition $U(0) = 0$ (i.e., initially zero energy in the lobe), is:
	\begin{equation}
		U(R) = \frac{(Q \Omega k c)^{1/2}}{[\alpha(\Gamma_c - 1)(q + 1) + (4 - \beta)/2] \kappa_1^{1/2}} R^{(4 - \beta)/2} .
	\end{equation}

	{The} average lobe pressure is, meanwhile, related to the total energy in the lobe cavity and its volume (i.e., Equation \eqref{udensity}, or, e.g., Equation (15) of \citet{KDA+1997}). We can, therefore, rewrite the above expression for the total energy in terms of the lobe pressure, 
	recalling $V(R) = \kappa_2 R^\alpha$, as:
	\begin{equation}
		p(R) = \frac{(Q \Omega k c)^{1/2} (\Gamma_c - 1)(q + 1)}{[\alpha(\Gamma_c - 1)(q + 1) + (4 - \beta)/2] \kappa_1^{1/2} \kappa_2 } R^{(4 - \beta - 2\alpha)/2} .
		\label{pressapp}
	\end{equation}
	{This} relationship is presented in Equation \eqref{pressse} of Section \ref{sec:The First Models} as a function of the source age upon the further substitution of Equation \eqref{scheuer}.
	
	\subsection{Lobe Volume}
	\label{app:Lobe volume}
	
	The sidewards expansion rate of the lobe is derived by equating the internal pressure to the ram pressure presented by the ambient medium as the lobe widens, i.e., $\rho v_\perp^2 = p(t)$, where the ambient gas density is reasonably approximated as $\rho \sim kR^{-\beta}$ for somewhat spherical lobes. As discussed in Section \ref{sec:The First Models}, the width of the lobe at some location $r$ along the jet axis is:
	\begin{equation}
		\begin{split}
			R_\perp(r) &= \int_{t(r)}^{t(R)} v_\perp(t^*) dt^* \\
			&= \int_{r}^{R} v_\perp(R^*) \frac{dt}{dR^*} dR^* ,
		\end{split}
	\end{equation}
	where $t(r)$ is the time when the jet head reached the location $r$ along the jet axis and $t(R)$ is the current time (i.e., jet head has length $R$). The change of variables in the second equality allows the width of the lobe to be evaluated in terms of known limits and the pressure in Equation \eqref{pressapp}. The integral is evaluated upon substitution of Equations \eqref{veloc} and \eqref{pressapp}:
	\begin{equation}
		\begin{split}
			R_\perp(r) &= \frac{4 (\Omega c)^{3/4} k^{1/4}}{(12 - \beta - 2\alpha) \kappa_1^{3/4} \kappa_2^{1/2} Q^{1/4}} \left[\frac{(\Gamma_c - 1)(q + 1)}{\alpha(\Gamma_c - 1)(q + 1) + (4 - \beta)/2} \right]^{1/2}\\
			&\quad\quad\quad\times( R^{(12 - \beta - 2\alpha)/4} - r^{(12 - \beta - 2\alpha)/4} ).
		\end{split}
	\end{equation}
	{The} lobe volume is derived in Equation \eqref{volsh} of Section \ref{sec:The First Models} by integrating this expression over all locations $r$ along the jet axis.

	\printendnotes[custom] 
	
	\begin{adjustwidth}{-\extralength}{0cm}
		
		\reftitle{References}
		
		
		

		\PublishersNote{}
		
		%
		
		
	\end{adjustwidth}
\end{document}